\documentclass[12pt]{article}
\usepackage{amsmath,amssymb,amsfonts,graphicx,bm}
\usepackage{pgfplots}
\usepackage{graphicx,bm}
\usepackage{geometry}
  
\renewcommand{\theequation}{\arabic{section}.\arabic{equation}}

\renewcommand{\theequation}{\arabic{section}.\arabic{equation}}

\newcommand{\calU}{{\mathcal U}}

\newcommand{\calP}{{\mathcal P}}
\newcommand{\calPP}{\widetilde{\mathcal P}}
\newcommand{\calQ}{{\mathcal Q}}
\newcommand{\calQQ}{\widetilde{\mathcal Q}}

\newcommand{\R}{{\mathbb R}}

\renewcommand{\P}{\mathbb{P}}

\newcommand{\x}{\mathbf{x}}

\newcommand{\e}{{\mathrm e}}
\newcommand{\E}{{\mathbb E}}

\newcommand{\calT}{{\mathcal T}}
\newcommand{\calS}{{\mathcal S}}
\newcommand{\calR}{{\mathcal R}}

\newcommand{\calM}{{\mathcal M}}

\renewcommand{\P}{\mathbb P}
\newcommand{\p}{\widetilde{p}}

\newcommand{\q}{\widetilde{q}}

\begin{document}

\title{Stochastic calculus of run-and-tumble motion: an applied perspective} 

\author{ \em
Paul. C. Bressloff, \\ Department of Mathematics, Imperial College London, \\
London SW7 2AZ, UK}

\maketitle

\begin{abstract}
The run-and-tumble particle (RTP) is one of the simplest examples of an active particle in which the direction of constant motion randomly switches. In the one-dimensional (1D) case this means switching between rightward and leftward velocities.
Most theoretical studies of RTPs are based on the analysis of the Chapman-Kolmogorov (CK) differential equation describing the evolution of the joint probability densities for particle position and velocity state. In this paper we develop an alternative, probabilistic framework of 1D RTP motion based on the stochastic calculus of Poisson and diffusion processes. In particular, we show how a generalisation of It\^o's lemma provides a direct link between sample paths of an RTP and the underlying CK equation. This allows us to incorporate various non-trivial extensions in a systematic fashion, including stochastic resetting and partially absorbing sticky boundaries. The velocity switching process and resetting process are represented by a pair of independent Poisson processes, whereas a sticky boundary is modelled using a boundary layer. We then use the probabilistic formulation to calculate stochastic entropy production along individual trajectories of an RTP, and show how the corresponding Gibbs-Shannon entropy is recovered by averaging over the ensemble of sample paths. Finally, we extend the probabilistic framework to a population of RTPs and use this to explore the effects of global resetting. 
 \end{abstract}

 \section{Introduction}  In recent years there has been considerable interest in {\em active matter}, which describes  ensembles of self-propelled elements that locally consume energy from the environment in order to move or to
exert mechanical forces \cite{Ramaswamy10,Palacci10,Vicsek12,Roman12,Bricard13,Attanasi14,Solon15,Cates15,Bechinger16}. Notable examples include motile bacteria, animal flocks or herds, and synthetically produced self-catalytic colloids. The persistent nature of active particle motion can lead to novel phenomena such as accumulation at walls and motility-based phase separation; the former occurs even in the absence of particle interactions. One of the simplest examples of an active particle is a run-and-tumble particle (RTP) moving in one spatial dimension. The particle randomly switches at a Poisson rate $\alpha$ between a left-moving ($\sigma =1$) and a right-moving ($\sigma =-1)$ constant velocity state $\sigma v$, $v>0$. An RTP model is motivated by the so-called `run-and-tumble' motion of bacteria such as {\em E. coli} \cite{Berg04}. In the presence of an attractive (repulsive) chemotactic signal, the switching rate becomes biased so that runs towards (away from) the signal source tend to persist. Let $X(t)$ and $\sigma(t)$ denote the position and velocity state of the RTP at time $t$. The dynamics can be described by the piecewise deterministic equation
\begin{equation}
\frac{dX(t)}{dt}=\sigma(t) v,
\label{RTP0}
\end{equation}
where $\sigma(t)\in \{-1,1\}$ evolves according to a two-state continuous-time Markov chain with transitions $\sigma(t) \overset{\alpha} \rightarrow -\sigma(t)$. (For simplicity, in this paper we focus on unbiased motion. However, one could consider biased motion in which $|v_{-1}|\neq v_1$ or the transitions $-1\rightarrow 1$ and $1\rightarrow -1$ have different rates.) Most theoretical studies of RTPs focus on the distribution of sample paths expressed as the solution to a forward or backward Chapman-Kolmogorov (CK) equation.
In the case of the standard model given by equation (\ref{RTP0}), the probability densities $p_k(x,t)$,
\[p_k(x,t)dx=\P[x\leq X(t)\leq x+dx,\sigma(t)=k],\]
satisfy the forward CK
\begin{align}
 \frac{\partial p_k(x,t)}{\partial t}=-vk\frac{\partial p_k(x,t)}{\partial x}+\alpha [p_{-k}(x,t)-p_k(x,t)],\quad k=\pm 1,\quad x \in \R.
\label{CK0}
\end{align}
There have been various extensions of the basic 1D model, including confined motion on the half-line or finite interval \cite{Angelani15,Angelani17,Malakar18,Bressloff22d,Bressloff23,Angelani23} and stochastic resetting \cite{Evans18,Bressloff20,Santra20a}. For example, suppose that an RTP is restricted to the half-line $[0,\infty)$ with $x=0$ treated as a hard wall where the particle becomes stuck until it reverses direction from $-v$ to $v$. (The rate of tumbling at the wall may differ from that in the bulk.) This is an example of a sticky boundary \cite{Angelani15,Angelani17,Bressloff23}. One way to model a sticky boundary is to introduce the probability $Q_0(t)$ that at time $t$ the particle is attached to the left-hand boundary in the bound state $B_0$. Equation (\ref{CK0}) is then supplemented by the boundary condition 
\begin{equation}
\label{mstick0}
\gamma Q_0(t)=vp_1(0,t) ,
\end{equation}
where $\gamma$ is the tumbling rate at the wall and $Q_0(t)$ evolves according to the equation
\begin{equation}
\label{stick0}
\frac{dQ_0}{dt}=vp_{-1}(0,t)-\gamma  Q_0(t) .
\end{equation}
Note that in the limit $\gamma \rightarrow \infty$, the particle returns to the bulk as soon as it hits the wall and we have the totally reflecting boundary condition
$
p_1(0,t)=p_{-1}(0,t)$.
On the other hand, taking the limit $\gamma \rightarrow 0$ leads to the totally absorbing boundary condition
$p_1(0,t)=0$.
The sticky boundary can be further modified to include the possibility of absorption whilst in the bound state $B_0$. If absorption occurs at a constant rate $\kappa$, then $\gamma \rightarrow \gamma+\kappa$ in equation (\ref{stick0}). 
As a second example, suppose that at a random sequence of times the state of an RTP is instantaneously reset to its initial state, that is,
$(X(t),\sigma(t))\rightarrow (x_0,\sigma_0)$,
where $X(0)=x_0$ and $\sigma(0)=\sigma_0$. (One could also consider a resetting protocol in which only the position $X(t)$ is reset \cite{Evans18}.) The sequence of resetting times is typically generated from a Poisson process with rate $r$. Stochastic resetting of the position and velocity state can be incorporated into the CK equation (\ref{CK0}) according to
\begin{align}
 \frac{\partial p_k(x,t)}{\partial t}=-vk\frac{\partial p_k(x,t)}{\partial x}+\alpha [p_{-k}(x,t)-p_k(x,t)] -rp_k(x,t)+r\delta(x-x_0)\delta_{k,\sigma_0} 
\label{CK0r}
\end{align}
for $k=\pm 1$.

In this paper we develop an alternative, probabilistic formulation of RTPs with stochastic resetting and/or sticky boundaries using the stochastic calculus of Poisson and diffusion processes. The velocity switching process and resetting process are represented by a pair of independent Poisson processes, whereas a sticky boundary is modelled using a boundary layer. We show how the probabilistic formulation has several advantages. First, there exists a generalised version of It\^o's lemma that provides a direct link between sample paths of an RTP and the underlying CK equation. More specifically, the generalised It\^o's lemma is used to derive a stochastic partial differential equation (SPDE) for an appropriately defined empirical measure. Averaging the SPDE with respect to sample paths then yields the corresponding CK equation. We apply this method to an RTP with diffusion in      
section 2, and an RTP with position and velocity resetting in section 3. Second, see section 4, 
the stochastic calculus of RTPs provides a powerful mathematical framework for incorporating encounter-based models of partial absorption at a sticky boundary. Analogous to encounter-based models of diffusion \cite{Grebenkov20,Grebenkov22,Bressloff22,Bressloff22a}, the RTP is absorbed as soon as the amount of contact time with the boundary (time spent in the boundary layer or bound state) crosses some randomly generated threshold. In our previous treatment of encounter-based models of RTPs, we used heuristic arguments to construct the appropriate CK equations \cite{Bressloff22d,Bressloff23}. Third, the stochastic calculus of RTPs can be used to evaluate stochastic entropy production along individual sample paths of an RTP (see section 5). Determining stochastic entropy production is of considerable interest within the context of stochastic thermodynamics, which extends classical ideas of entropy, heat and work to mesoscopic non-equilibrium systems. \cite{Seifert05,Sekimoto10,Seifert12,Cocconi20,Peliti21,Roldan23}. Although stochastic entropy production does not satisfy the second law of thermodynamic, the latter is recovered after averaging the stochastic entropy over the ensemble of stochastic trajectories to yield the corresponding Gibbs-Shannon entropy, which was recently derived for RTPs in Refs. \cite{Cocconi20,Frydel22,Angelani24}. Finally, in section 6 we highlight another useful feature of the probabilistic formulation, namely, it can be used to derive an SPDE for the global density of a population of RTPs. We illustrate the theory by considering the effects of global resetting, whereby all particles are simultaneously reset to their initial states. 
This induces statistical correlations between the particle, even in the absence of particle-particle interactions. In addition, taking moments of the empirical measure leads to a hierarchy of moment equations that takes the form of differential equations with resetting in which lower-order moments are embedded into the equations at higher orders..

\setcounter{equation}{0}
\section{RTP with diffusion in $\R$}

Consider an RTP that randomly switches at a rate $\alpha$ between two constant velocity state labeled by $v_{\pm 1}$ with $v_1=v$ and $v_{-1}=-v$ for some $v>0$. Let $X(t)$ and $\sigma(t)$ denote the position and direction of the RTP at time $t$ with $X(0)=x_0$ and $\sigma(0)=\sigma_0$. In between velocity reversal events, the particle evolves according to the SDE 
\begin{equation}
\label{dX}
dX(t)=v\sigma(t)dt +\sqrt{2D}dW(t),
 \end{equation}
 where $D$ is the diffusivity and $W(t)$ is a Brownian motion. The corresponding integral version is
 \begin{equation}
\label{intX}
X(t)=x_0+v\int_0^t \sigma(s)ds +\sqrt{2D}\int_0^tdW(s),
 \end{equation}
 
 \subsection*{Stochastic calculus of jump processes} 
 
\begin{figure}[b!]
\centering
\includegraphics[width=8cm]{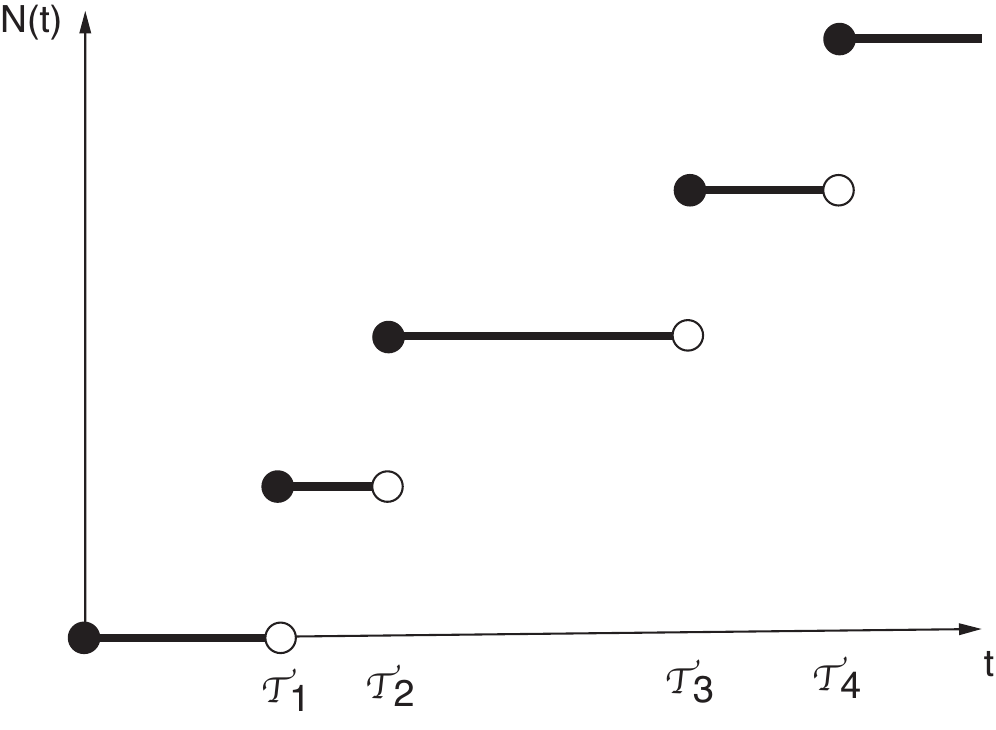}
\caption{One path of a Poisson process $N(t)$, illustrating that it is right-continuous. Jumps occur at the times $\calT_{\ell}$, $\ell \geq 1$.}
\label{fig1}
\end{figure}

In order to develop the stochastic calculus of RTPs, we combine the SDE (\ref{dX}) with a corresponding equation for the differential $d\sigma(t)$. First, we need to specify the times at which direction reversals occur during a particular realisation of the dynamics for $(X(t),\sigma(t))$. Let $\calT_n$, $n \geq 1$, denote the $n$th time that the RTP reverses direction and introduce the inter-reversal times $\tau_n=\calT_{n}-\calT_{n-1}$ with
\begin{equation}
\P[\tau_n\in [s,s+ds]]=\alpha \e^{-\alpha s}ds.
\end{equation} 
 The number of direction reversals occurring in the time interval $[0,t]$ is given by the Poisson process
 $N(t)$ with
 \begin{align}
N(t)=n,\quad \calT_{n}\leq t <\calT_{n+1}.
\end{align}
Note that $N(t)$ is defined to be right-continuous. That is, $N (\calT_n^-)=n-1$ whereas $N(\calT_n)=n$. This is illustrated in Fig. \ref{fig1}.
The probability distribution of the Poisson process is
\begin{align}
\P[N(t)=n]=  \frac{(\alpha t)^{n}\e^{-\alpha t}}{n!},
\label{Pois1}
\end{align}
with $\E[N(t)]=\alpha t=\mbox{Var}[N(t)]$.

Suppose that $\sigma(t)=\sigma_n$ for $\calT_n \leq t < \calT_{n-1}$ with $\calT_0=0$. Using telescoping,  we write the jump process in the form
\begin{align}
  \sigma(t)-\sigma_0
&=\sum_{n=1}^{N(t)}[\sigma_n-\sigma_{n-1}]=\sum_{n=1}^{N(t)}[\sigma(\calT_n)-\sigma(\calT_{n-1})] =\sum_{n=1}^{N(t)}[\sigma(\calT_n)-\sigma(\calT_{n}^-)] .
\end{align}
Since each jump involves a sign reversal, it follows that
\begin{align}
 \sigma(t)&=\sigma_0-2\sum_{n=1}^{N(t)}\sigma(\calT_{n}^-)=\sigma_0-2\int_0^t \sigma(s^-)dN(s)
\label{intsig}
\end{align}
with
\begin{equation}
 dN(t)=h(t)dt,\quad h(t)=\sum_{n=1}^{\infty} \delta(t-\calT_n).
\end{equation}
Equation (\ref{intsig}) is the discrete analog of equation (\ref{intX}).
Formally differentiating both sides of equation (\ref{intsig}) then yields the discrete analog of equation (\ref{dX}), namely
\begin{align}
d\sigma(t)&= -2  \sigma(t^-)dN(t)
\label{dsig}
\end{align}
The advantage of this formulation is that it can be used to derive a generalised It\^o's lemma.

 \subsection*{Generalised It\^o's lemma}
 
 Let $f$ be an arbitrary smooth bounded function on $\R\times \{-1,1\}$.  Combining It\^o's  integral formula for SDEs with the jump process for $\sigma(t)$ gives
\begin{align}
  f(X(t),\sigma(t))
&=f(x_0,\sigma_0)+\int_0^t \bigg [v\sigma(s) f'(X(s),\sigma(s))+D  f''(X(s),\sigma(s)) \bigg ]ds\\
  &\ +\sqrt{2D} \int_0^tf'(X(s),\sigma(s)) dW(s) \nonumber \\
  &\quad +\sum_{n=1}^{N(t)} \bigg [f(X(\calT_n),-\sigma(\calT_n^-)))-f(X(\calT_n),\sigma(\calT_n^-))\bigg ].\nonumber 
 \end{align}
 where $'$ denotes differentiation with respect to $X$. In terms of the Poison process $N(t)$, this equation can be rewritten as the generalised It\^o integral formula
 \begin{align}
  f(X(t),\sigma(t))
&=f(x_0,\sigma_0)+\int_0^t \bigg [v\sigma(s) f'(X(s),\sigma(s))+D  f''(X(s),\sigma(s)) \bigg ]ds \\
  &\ +\sqrt{2D} \int_0^tf'(X(s),\sigma(s)) dW(s) \nonumber \\&\quad  +\int_0^t\bigg [f(X(s),-\sigma(s^-))-f(X(s),\sigma(s^-))\bigg ]dN(s).\nonumber \end{align}
The differential version is thus
\begin{align}
\label{lemma0}
 df(X(t),\sigma(t))&=\bigg [v\sigma(t) f'(X(t),\sigma(t))+D  f''(X(t),\sigma(t)) \bigg ]dt \\
 &\quad +\sqrt{2D} f'(X(t),\sigma(t)) dW(t)  \nonumber \\
 &\quad +\bigg [f(X(t^-),-\sigma(t^-))-f(X(t^-),\sigma(t^-))\bigg ]dN(t).\nonumber
\end{align}

Throughout this paper we will take expectations with respect to white noise and Poisson processes. The former can be carried out using the standard property of It\^o calculus, namely, $dW(t)=W(t+dt)-W(t)$ is statistically independent of the current position $X(t)$. An analogous property holds for Poisson jump processes. More specifically, given some measurable function $F(x,\sigma)$, we have 
\begin{equation}
\E[F(X(t^-),\sigma(t^-))dN(t)]=\E[F(X(t^-),\sigma(t^-))]\E[dN(t)],
\end{equation}
since $F(X(t),\sigma(t))$ for all $t<\calT_n$ only depends on previous jump times. Moreover,
$N(t)-N(\tau)=\int_{\tau}^tdN(s)$ so that $\int_{\tau}^t\E[dN(s)] = \E[N(t)-N(\tau)]=\alpha (t-\tau)$ and, hence,
$\E[dN(t)]=\alpha dt$.

 \subsection*{Derivation of the Chapman-Kolmogorov equation}
 One immediate application of  the generalised It\^o formula (\ref{lemma0}) is to derive the corresponding differential CK equation for the SDE (\ref{dX}).
Introducing the empirical measures
\begin{equation}
\rho_{k}(x,t)=\delta(x-X(t))\delta_{k,\sigma(t)},\quad k=\pm 1,
\label{meas}
\end{equation}
we have the identity
\begin{equation}
\sum_{k=\pm 1} \int_{\R}\rho_k(x,t)f(x,k)dx=f(X(t),\sigma(t)).
\end{equation}
Taking differentials of both sides with respect to $t$ yields
\begin{align}
\label{ff0}
df(t)=\left [ \sum_{k=\pm 1} \int_{\R} f(x,k)\frac{\partial \rho_k(x,t)}{\partial t}  dx\right ]dt  
 \end{align}
 Substituting equation (\ref{lemma0}) into the left-hand side of equation (\ref{ff0}) and using the definition of $\rho_k$ gives
    \begin{align}
 & \sum_{k=\pm 1} \int_{\R}   f(x,k)\frac{\partial \rho_k(x,t)}{\partial t}  dx \nonumber \\
  &\quad =\sum_{k=\pm 1}\int_{\R} \rho_k(x,t) \bigg[ vk f'(x,k)+D  f''(x,k)+\sqrt{2D} f'(x,k)\xi(t) \bigg]dx \nonumber\\
   &\quad +h(t) \sum_{k=\pm 1}  \int_{\R}  [\rho_k(x,t^-)f(x,-k)-\rho_k(x,t^-)f(x,k)]dx . 
    \label{step}
    \end{align}
    We have formally set $dW(t)=\xi(t)dt$ where $\xi(t)$ is white noise process satisfying $\langle \xi(t)\rangle =0$ and $\langle \xi(t)\xi(t')\rangle =\delta(t-t')$.
 Performing an integration by parts and resumming yields
   \begin{align}
 &\sum_{k=\pm 1} \int_{\R}   f(x,k)\frac{\partial \rho_k(x,t)}{\partial t}  dx \nonumber \\
 &\quad =\sum_{k=\pm 1}\int_{\R} f(x,k) \bigg[ -vk \rho_k'(x,t)+D  f''(x,k)+\sqrt{2D} \rho_k'(x,t)\xi(t) \bigg]dx \nonumber\\
  &\quad +h(t) \sum_{k=\pm 1}  \int_{\R} f(x,k)[\rho_{-k}(x,t^-)-\rho_{k}(x,t^-)]dx .\end{align}
Since the functions $f(x,k)$ are arbitrary, we obtain the stochastic partial differential equation (SPDE)
\begin{align}
  \frac{\partial \rho_k}{\partial t}&= -vk\frac{\partial \rho_k}{\partial x} +D\frac{\partial^2 \rho_k}{\partial x^2}+\sqrt{2D}\frac{\partial \rho_k}{\partial x}\xi(t)
  +h(t) [\rho_{-k}(x,t^-)-\rho_{k}(x,t^-)].
 \label{SPDE}
\end{align}
Finally, define the indexed set of probability densities
\begin{equation}
p_k(x,t)=\left \langle \E \left [\rho_k(x,t)\right ]\right \rangle\equiv \left \langle \E \left [\delta(x-X(t))\delta_{k,\sigma(t)} \right ]\right \rangle ,
\label{defpk}
\end{equation}
where $\langle \cdot \rangle$ and $\E[\cdot]$ denote expectations with respect to the white noise process and the Markov chain, respectively. We then take expectations of equation (\ref{SPDE}) with respect to both noise processes using the identities 
\begin{align}
\left \langle \E \left [\rho_k(x,t^-)h(t)\right ]\right \rangle= \left \langle \E \left [\rho_k(x,t^-)\right ]\right \rangle \E \left [ h(t)\right ],\quad  \E \left [ h(t)\right ]=\alpha,
\end{align}
This leads to the CK equation
\begin{align}
 \frac{\partial p_k(x,t)}{\partial t}=-vk\frac{\partial p_k(x,t)}{\partial x}+D\frac{\partial^2 p_k(x,t)}{\partial x^2}+\alpha [p_{-k}(x,t)-p_k(x,t)],\quad k=\pm 1,
\label{CK}
\end{align}
which recovers equation (\ref{CK0}) when $D=0$.
We summarise the steps in the derivation of the CK equation as follows.
\medskip

\noindent 1. Write down the underlying SDE with jumps in terms of one or more Poisson processes
\medskip

\noindent 2. Construct a version of It\^o's lemma for an arbitrary test function
\medskip

\noindent 3. Use the generalised It\^o's lemma and integration by parts to derive an SPDE for the corresponding empirical measure
\medskip

\noindent 4. Average the SPDE with respect to the various independent noise sources to obtain the CK equation.

\setcounter{equation}{0}
\section{RTP with stochastic resetting}

As our first extension of the mathematical approach presented in section 2, we consider run-and-tumble motion with instantaneous stochastic resetting. Suppose that the position $X(t)$ and velocity state $\sigma(t)$ of an RTP reset to their initial values $x_0$ and $\sigma_0$, respectively, at a constant rate $r$. There are now two independent jump processes, one associated with velocity reversal events and the other with resetting events. Following section 2, let $\calT_n$, $n \geq 1$, denote the $n$th time that the RTP reverses direction, with the number of jumps in the time interval $[0,t]$ given by the Poisson process $N(t)$, with probability distribution (\ref{Pois1}). Similarly, let $\overline{\calT}_n$ , $n\geq 1$, denote the $n$th time that the RTP resets. The inter reset times $\overline{\tau}_n =\overline{\calT}_n-\overline{\calT}_{n-1}$ are exponentially distributed with
\begin{equation}
 \P[\overline{\tau}_n\in [s,s+ds]]=r \e^{-r s}ds
\end{equation} 
 In addition, the number of resets occurring in the time interval $[0,t]$ is given  by the Poisson process 
  $\overline{N}(t)$ with
 \begin{align}
\overline{N}(t)=n,\quad \overline{\calT}_{n}\leq t <\overline{\calT}_{n+1},\quad \P[\overline{N}(t)=n]=  \frac{(r t)^{n}\e^{-r t}}{n!}.
\label{Pois2}
\end{align}

Since both $X(t)$ and $\sigma(t)$ reset, it is necessary to modify equations (\ref{dX}) and (\ref{dsig}). Jumps in $X(t)$ only occur due to resetting. As shown elsewhere \cite{Bressloff24a}, positional resetting can be implemented as a jump-diffusion process with
\begin{align}
 X(t)&=x_0+\int_0^tv\sigma(s)ds+2D\int_0^t dW(s)+\sum_{n=1}^{\overline{N}(t)}  [x_0-X(\overline{\calT}_n^-)].
\label{intXr}
\end{align}
Setting $t=\overline{\calT}_n$ and $t=\overline{\calT}_n^-$ in equation (\ref{intXr}) and subtracting the resulting pair of equations shows that $X(\overline{\calT}_n)= X(\overline{\calT}_{n}^-)+x_{0}- X(\overline{\calT}_{n}^-)=x_0$
which represents to instantaneous resetting. The corresponding SDE is thus
\begin{equation}
dX(t)=v\sigma(t)dt+\sqrt{2D}dW(t)+(x_0-X(t^-))d\overline{N}(t),
\label{dXr}
\end{equation}
where
\begin{align}
\label{dNdefr}
d\overline{N}(t) =\overline{h}(t)dt,\quad \overline{h}(t)=\sum_{k=1}^{\infty}\delta(t-\overline{\calT}_{k})  .
\end{align}
In contrast to $X(t)$, the discrete state $\sigma(t)$ jumps at both direction reversal and resetting events.
Let $\calS_{m}$, $m \geq 1$, denote the resulting sequence of jump times so that
\[\{\calS_1,\ldots ,\calS_{N_{\rm tot}(t)}\}=\{\calT_1,\ldots ,\calT_{N(t)}\}\cup \{\overline{\calT}_1,\ldots ,\overline{\calT}_{\overline{N}(t)}\}\]
with $N_{\rm tot}(t)=N(t)+\overline{N}(t)$. (Note that resetting only produces a jump in $\sigma(t)$ following an odd number of switching events.) Setting $\sigma(t)=\sigma_m$ for $\calS_m\leq t < \calS_{m+1}$ with $\calS_0=0$, we use telescoping to write 
\begin{align}
  \sigma(t)-\sigma_0
&=\sum_{m=1}^{N_{\rm tot}(t)}[\sigma_m-\sigma_{m-1}]\nonumber \\
&=\sum_{m=1}^{N_{\rm tot}(t)}[\sigma(\calS_m)-\sigma(\calS_{m-1})] =\sum_{m=1}^{N_{\rm tot}(t)}[\sigma(\calS_m)-\sigma(\calS_{m}^-)] .
\end{align}
We now partition the jumps into resetting and direction reversal events. In the former case we have $\sigma(\calS_m)=\sigma_0$, whereas in the latter case $\sigma(\calS_{m})=-\sigma(\calS_{m}^-)$. Hence,
\begin{align}
 \sigma(t)&=\sigma_0
+\sum_{n=1}^{\overline{N}(t)}[\sigma_0-\sigma(\overline{\calT}_n^-)]-2\sum_{n=1}^{N(t)}\sigma(\calT_{n}^-),
\end{align}
which has the equivalent integral form
\begin{align}
\sigma(t)
= \sigma_0 +\int_0^t [\sigma_0-\sigma(s^-)]d\overline{N}(s)-2\int_0^t \sigma(s^-)dN(s),
\label{intsigr}
\end{align}
The corresponding differential version of this equation is
\begin{align}
d\sigma(t)&=  [\sigma_0-\sigma(t^-)]d\overline{N}(t)-2  \sigma(t^-)dN(t).
\label{dsigr}
\end{align}

Following along similar lines to section 2, we can use equations (\ref{dXr}) and (\ref{dsigr}) to write down the generalised It\^o's lemma for an RTP with diffusion and stochastic resetting:
\begin{align}
 df(X(t),\sigma(t))&=\bigg [v\sigma(t) f'(X(t),\sigma(t))+D  f''(X(t),\sigma(t)) \bigg ]dt\nonumber \\
 &\quad +\sqrt{2D} f'(X(t),\sigma(t)) dW(t) +\bigg [f(x_0,\sigma_0)-f(X(t^-),\sigma(t^-))\bigg ]d\overline{N}(t)\nonumber \\
 &\quad +\bigg [f(X(t),-\sigma(t^-))-f(X(t),\sigma(t^-))\bigg ]dN(t).\label{lemmar}
\end{align}
This, in turn, implies that equation (\ref{step}) for the empirical measure $\rho_k(x,t)$ picks up the additional term     \begin{align}
\overline{h}(t) \sum_{k=\pm 1}  \int_{\R}  [\delta(x-x_0)\delta_{k,\sigma_0}-\rho_k(x,t^-)]f(x,k)]dx.
\end{align}
 Performing an integration by parts and using the arbitrariness of $f(x,k)$ then yields the modified
  SPDE
\begin{align}
  \frac{\partial \rho_k}{\partial t}&=D\frac{\partial^2 \rho_k}{\partial x^2} -vk\frac{\partial}{\partial x} \rho_k+\sqrt{2D}\frac{\partial \rho_k}{\partial x}\xi(t)\nonumber \\
  &\quad   +\overline{h}(t)  [\delta(x-x_0)\delta_{k,\sigma_0}-\rho_k(x,t^-)] 
  +h(t) [\rho_{-k}(x,t^-)-\rho_k(x,t^-)].
 \label{SPDEH}
\end{align}
Finally, averaging with respect to the white noise process and the two jump processes,
we obtain the CK equation for an RTP with diffusion and resetting:
\begin{equation}
\label{CKH}
  \frac{\partial p_k}{\partial t}=-\frac{\partial J_k(x,t)}{\partial x}+r\delta(x-x_0)\delta_{k,\sigma_0}-rp_k(x,t)+\alpha[p_{-k}(x,t)-p_k(x,t)]
\end{equation}
with 
\begin{equation}
\label{Jk}
J_k(x,t)=kvp_k(x,t)-D\frac{\partial p_k(x,t)}{\partial x}.
\end{equation}
Equation (\ref{CKH}) reduces to (\ref{CK0r}) when $D=0$.

\setcounter{equation}{0}
\section{RTP in $\R^+$ with a sticky boundary at $x=0$}

As our second extension of the analysis in section 2, we consider an RTP confined to the half-line $[0,\infty)$ with a sticky boundary at $x=0$ that is either non-absorbing or partially absorbing. Since dealing with sticky boundary conditions for diffusion processes is considerably more involved, we set $D=0$ throughout this section. For the moment, we also assume that there is no resetting.

\subsection{Non-absorbing sticky boundary at $x=0$.}

\begin{figure}[t!]
\centering
\includegraphics[width=12cm]{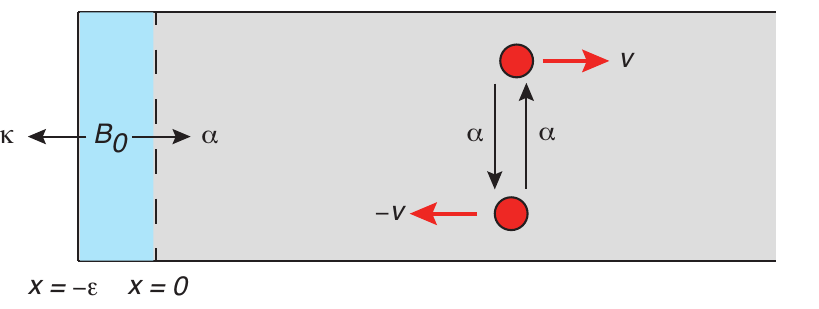}
\caption{Schematic representation of an RTP confined to the domain $[0,\infty)$ with a sticky boundary at one end. We represent the latter as a boundary layer $B_0$ of width $\epsilon$. A left-moving particle at $x=0^+$ enters the boundary layer where it remains until reversing direction and exiting at a rate $\alpha$. Alternatively, the bound particle may be absorbed at a rate $\kappa$. In the absence of absorption we have $\kappa=0$.}
\label{fig2}
\end{figure}

First suppose that the boundary at $x=0$ represents a hard wall, which is an example of a sticky boundary without absorption \cite{Angelani15,Angelani17,Bressloff23}. We incorporate the boundary condition by introducing a thin boundary layer $[-\epsilon,0]$ and taking $X(t)\in \Omega_{\epsilon}\equiv [-\epsilon ,\infty)$, see Fig. \ref{fig2}. When the RTP reaches $x=0^+$ in the left-moving state it enters the boundary layer, where it remains until reversing direction at a rate $\alpha$. (In order to simplify the notation, we take the tumbling rate in the boundary layer to be the same as in the bulk.) It then re-enters the bulk domain in the right-moving state. This can be represented by the sequence of transitions
\begin{equation}
(0^+,-1)\rightarrow (0^-,-1) \rightarrow  (0^+,1).
\end{equation}
The analogs of equations (\ref{dX}) and (\ref{dsig}) are then
\begin{align}
\label{dXsticky0}
  dX(t)&=v\sigma(t)dt{\bf 1}_{X(t)>0},\quad  d\sigma(t)= -2  \sigma(t^-)dN(t)
  \end{align}
  for all $ X(t)\in \Omega_{\epsilon}$.

 Let $f$ be an arbitrary smooth bounded function on $\Omega_{\epsilon} \times \{-1,1\}$.  Applying It\^o's formula for  jump processes we have
\begin{align}
 df(X(t),\sigma(t))&= v\sigma(s) f'(X(t),\sigma(t)){\bf 1}_{X(t)> 0}  dt\nonumber \\
 &\quad +\bigg [f(X(t),-\sigma(t^-))-f(X(t),\sigma(t^-))\bigg ]\, dN(t).
\label{lemma2}
\end{align}
Introducing the empirical measures
\begin{equation}
 \rho_{\epsilon,k}(x,\ell,t)=\delta(x-X(t))\delta_{k,\sigma(t)},\ X(t)\in \Omega_{\epsilon},\ k=\pm 1,\end{equation}
we have the identity
\begin{equation}
 \sum_{k=\pm 1} \int_{-\epsilon}^{\infty}\  \rho_{\epsilon,k}(x,t)f(x,k)dx=f(X(t),\sigma(t)).
\end{equation}
Taking differentials of both sides with respect to $t$ yields
\begin{align}
\label{ff}
 \sum_{k=\pm 1} \int_{-\epsilon}^{\infty} f(x,k)\frac{\partial \rho_{\epsilon,k}(x,t)}{\partial t}  dx =\frac{df}{dt}(X(t),\sigma(t)).
 \end{align}
 From the generalised It\^o's lemma,    
 \begin{align}
    \sum_{k=\pm 1} \int_{-\epsilon}^{\infty} f(x,k)\frac{\partial \rho_{\epsilon,k}(x,t)}{\partial t}  dx
  &=\sum_{k=\pm 1} \int_{0} ^{\infty}\rho_{\epsilon,k}(x,t)  vk f'(x,k) dx \\
    &   +h(t)\sum_{k=\pm 1}   \int_{-\epsilon}^{\infty} [\rho_{\epsilon,k}(x,t^-)f(x,-k))-\rho_{\epsilon,k}(x,t^-)f(x,k)]dx .\nonumber
\end{align}
 Performing an integration by parts, we have
   \begin{align}
& \sum_{k=\pm 1} \left  [\int_{0}^{\infty} f(x,k)\frac{\partial \rho_{\epsilon,k}(x,t)}{\partial t}  dx+ \int_{-\epsilon}^{0} f(x,k)\frac{\partial \rho_{\epsilon,k}(x,t)}{\partial t}  dx  \right ]\nonumber \\
 &=-v\int_{0} ^{\infty}f(x,1)  \rho_{\epsilon,1}'(x,t)dx +v\int_{0} ^{\infty}f(x,-1)  \rho_{\epsilon,-1}'(x,t)dx+v\rho_{\epsilon, -1}(0^+,t)f(0^+,-1) \nonumber\\
 &\quad -v\rho_{\epsilon,1}(0^+,t)f(0^+,1)+h(t) \sum_{k=\pm 1}  \int_{-\epsilon}^{\infty} f(x,k)[\rho_{\epsilon, -k}(x,t^-)-\rho_{\epsilon,k}(x,t^-)]dx .
\label{reps}
\end{align}

The crucial step is to note that the probability density $\rho_{\epsilon,-1}(x,t)$ has non-zero measure $q(t)$ in the boundary layer so we set
\begin{equation}
\label{dec}
\rho_{\epsilon,k}(x,t)=\rho_k(x,t){\bf 1}_{x>0}+\delta(x+\epsilon)\delta_{k,-1}q(t),
\end{equation}
Substituting this decomposition into equation (\ref{reps}) and taking the limit $\epsilon \rightarrow 0$ gives
   \begin{align}
& \sum_{k=\pm 1} \int_{0}^{\infty} f(x,k)\frac{\partial \rho_{k}(x,t)}{\partial t}  dx+ f(0^-,-1)\frac{dq(t)}{dt}    \nonumber \\
 &=-v\int_{0} ^{\infty}f(x,1)  \rho_{1}'(x,t)dx +v\int_{0} ^{\infty}f(x,-1)  \rho_{-1}'(x,t)dx\nonumber \\
&\quad +v\rho_{ -1}(0^+,t)f(0^+,-1)-v\rho_{1}(0^+,t)f(0^+,1) \nonumber\\
  &\quad +h(t) \sum_{k=\pm 1}  \int_{0}^{\infty} f(x,k)[\rho_{-k}(x,t^-)-\rho_k(x,t^-)]dx  
+h(t)q(t)[f(0^-,1) -f(0^-,-1) ].
\end{align}
Imposing continuity of the functions $f(x,k)$ across $x=0$ and using the fact that they are otherwise arbitrary gives the SPDE
\begin{subequations}
\begin{align}
  \frac{\partial \rho_k}{\partial t}&= -vk\frac{\partial}{\partial x} \rho_k(x,t)+h(t) [\rho_{-k}(x,t^-)-\rho_k(x,t^-)],\ k=\pm 1,\ x >0,\\
   0&=h(t)q(t)-v\rho_1(0^+,t) ,\\
   \frac{dq}{d t}&=-h(t) q(t)+v\rho_{-1}(0^+,t).
 \label{SPDEsick}
\end{align}
\end{subequations}
Finally, taking expectations with respect to the Poisson process $N(t)$ and setting
\begin{equation}
p_k(x,t)=\E[\rho_k(x,t)],\quad Q(t)=\E[q(t)],
\end{equation}
we obtain the CK equation for an RTP in $\R^+$ with a hard wall at $x=0$:
\begin{subequations}
\begin{align}
\label{CKsticka}
 \frac{\partial p_k(x,t)}{\partial t}&=-vkp_k(x,t)+\alpha [p_{-k}(x,t)-p_k(x,t)],\quad k=\pm 1,\quad x \in \R,\\
\label{CKstickb}
 \alpha Q(t)&=vp_1(0^+,t),\\
  \frac{dQ(t)}{dt}&=-\alpha Q(t)+vp_{-1}(0^+,t),
\label{CKstickc}
\end{align}
\end{subequations}

\subsection{Partially absorbing sticky boundary at $x=0$.}

Now suppose that whenever the RTP is in the boundary layer (stuck to the wall) it can be absorbed before reversing direction and re-entering the bulk domain. That is, there is a partially absorbing sticky boundary. This scenario is illustrated in Fig. \ref{fig2} for a constant rate of absorption $\kappa>0$. In Ref. \cite{Bressloff23} we analyzed the resulting first passage time (FTP) problem for a more general class of partially absorbing sticky boundaries by adapting the encounter-based model of diffusion-mediated surface absorption \cite{Grebenkov20,Grebenkov22,Bressloff22,Bressloff22a}. In contrast to the above boundary layer formulation, we represented the sticky boundary as a bound state $B_0$ along the lines of equations (\ref{CK0})--(\ref{stick0}). 
First, we assumed that if the RTP is in the bound state at time $t$, then there is a non-zero probability of absorption that depends on the cumulative amount of time (or occupation time) $A(t)$ that the RTP has spent in the bound state over the time interval $[0,t]$. Second, we introduced the joint probability density or occupation time propagator
\begin{align}
 \calP_{k}(x,a,t)dxda =\P[x\leq X(t)\leq x+dx,a\leq A(t)\leq a+da,\sigma(t)=k],
\end{align}
with $X(0)=x_0$, $\sigma(0)=\sigma_0$ and $A(0)=0$. Similarly, we defined $Q(a,t)da$ to be the probability that the RTP is in the bound state at time $t$ with $a\leq A(t)\leq a+da$. We then wrote down the CK equation for the propagator based on a generalisation of equations (\ref{CK0})--(\ref{stick0}), using a heuristic argument to justify the modified boundary conditions. Finally, we introduced the stopping time condition
\begin{equation}
\label{TA}
{\mathcal T}=\inf\{t>0:\ A(t) >\widehat{A}\},
\end{equation}
where $\widehat{A}$ is a random variable with probability distribution $\P[\widehat{A}>a]=\Psi(a)$. Note that ${\mathcal T}$ is a random variable that specifies the first absorption time when the RTP is in a bound state, which is identified with the event that $A(t)$ first crosses a randomly generated threshold $\widehat{A}$. Given that $A(t)$ is a nondecreasing process, the condition $t < {\mathcal T}$ is equivalent to the condition $A(t)<\widehat{A}$. This implies that the marginal probability density for particle position $X(t) $ and for being in the bound state are \cite{Bressloff23}
\begin{align}
\label{peep}
   p^{\Psi}(x,t)&=\int_0^{\infty}\Psi(a) \calP(x,a,t)da  , \quad Q^{\Psi}(t)=\int_0^{\infty}\Psi(a) \calQ(a,t)da .
\end{align}
The case of a constant rate of absorption $\kappa$ is recovered by taking $\Psi(a)=\e^{-\kappa a}$.

In this section we derive the propagator CK equation from first principles using stochastic calculus and the boundary layer representation shown in Fig. \ref{fig2}. First, the SDE (\ref{dXsticky0}) is supplemented by an equation for the occupation time $A(t)$:
\begin{align}
  dX(t)&=v\sigma(t){\bf 1}_{X(t)>0}dt,\quad dA(t)=\delta_{\sigma(t) ,-1}{\bf 1}_{X(t)<0}dt\nonumber \\
  d\sigma(t)&= -2  \sigma(t^-)dN(t),
 \end{align}
 Let $f$ be an arbitrary smooth bounded function on $\Omega_{\epsilon} \times \R^+\times \{-1,1\}$.  Applying It\^o's formula for  jump processes we have
\begin{align}
 & df(X(t),A(t),\sigma(t))\nonumber \\
 & \quad = v\sigma(s) \partial_xf(X(t),A(t))\sigma(t)){\bf 1}_{X(t)> 0}  dt + \partial_af(X(t),A(t),\sigma(t) )\delta_{\sigma(t) ,-1}{\bf 1}_{X(t)<0}dt\nonumber \\
 &\qquad +\bigg [f(X(t),A(t),-\sigma(t^-))-f(X(t),A(t),\sigma(t^-))\bigg ]\, dN(t).\label{lemma3}
\end{align}
Introducing the empirical measures
\begin{equation}
  \rho_{\epsilon,k}(x,a,t)=\delta(x-X(t))\delta(a-A(t))\delta_{k,\sigma(t)}
\end{equation}
for $X(t)\in \Omega_{\epsilon},\ A(t)\in \R^+,\ k=\pm 1$, we have the identity
\begin{equation}
  \sum_{k=\pm 1} \int_{-\epsilon}^{\infty}\left [\int_0^{\infty}  \rho_{\epsilon,k}(x,a,t)f(x,a,k)da\right ]dx=f(X(t),A(t),\sigma(t)).
\end{equation}
Taking differentials of both sides with respect to $t$ yields
\begin{align}
\label{ffa}
  \sum_{k=\pm 1} \int_{-\epsilon}^{\infty}\left [ f(x,a,k)\frac{\partial \rho_{\epsilon,k}(x,a,t)}{\partial t}da\right ]  dx =\frac{df}{dt}(X(t),A(t),\sigma(t)).
 \end{align}
Using the generalised It\^o's lemma (\ref{lemma3}) and performing an integration by parts along analogous lines to the derivation  of equation (\ref{reps}) gives
   \begin{align}
 & \sum_{k=\pm 1}  \int_{-\epsilon}^{\infty} \left [ \int_0^{\infty} f(x,a,k)\frac{\partial \rho_{\epsilon,k}(x,a,t)}{\partial t}  da \right ]dx \nonumber \\
  &=v\int_{0} ^{\infty}\left \{\int_0^{\infty}\bigg [f(x,a,-1)  \partial_x\rho_{\epsilon,-1}(x,a,t)- f(x,a,1) \partial_x \rho_{\epsilon,1}(x,a,t) \bigg ]da \right \}dx \nonumber \\
  &\quad +v\int_0^{\infty} \bigg [\rho_{\epsilon, -1}(0^+,a,t)f(0^+,a,-1)-\rho_{\epsilon,1}(0^+,a,t)f(0^+,a,1)\bigg ]da \nonumber\\
  & \quad - \int_{-\epsilon}^0 \left [\int_0^{\infty} f(x,a,-1)\partial_a\rho_{\epsilon,-1}(x,a,t)da \right ]dx- \int_{-\epsilon}^0  f(x,0,-1)\rho_{\epsilon,-1}(x,0,t)dx\nonumber \\
  &\quad +h(t) \sum_{k=\pm 1}  \int_{-\epsilon}^{\infty} \left \{\int_0^{\infty}  f(x,a,k)\bigg [\rho_{\epsilon,-k}(x,a,t^-)-\rho_{\epsilon,k}(x,a,t^-) \bigg ]da \right \}dx .
\label{reps2}
\end{align}

As in the non-absorbing case, we decompose the density as
\begin{equation}
\rho_{\epsilon,k}(x,a,t)=\rho_k(x,a,t){\bf 1}_{x>0}+\delta(x+\epsilon)\delta_{k,-1}q(a,t),
\end{equation}
Substituting this decomposition into equation (\ref{reps2}) and taking the limit $\epsilon \rightarrow 0$ gives
   \begin{align}
 & \sum_{k=\pm 1} \int_{0}^{\infty} f(x,a,k)\frac{\partial \rho_{k}(x,a,t)}{\partial t}  dx+ f(0^-,a,-1)\frac{\partial q(a,t)}{\partial t}    \nonumber \\
  &=v\int_{0} ^{\infty}\left \{\int_0^{\infty} \left [f(x,a,-1)  \partial_x\rho_{-1}(x,a,t)-f(x,a,1)  \partial_x \rho_{1}(x,a,t)\right ]da \right \}dx\nonumber \\
 &\quad +v\int_0^{\infty} \bigg [\rho_{ -1}(0^+,a,t)f(0^+,a,-1)-v\rho_{1}(0^+,a,t)f(0^+,a,1)\bigg ] da\nonumber\\
 & \quad - \int_0^{\infty} f(0^-,a,-1)\partial_aq(a,t)da -   f(0^-,0,-1)q(0,t)\nonumber \\
   &\quad +h(t) \sum_{k=\pm 1}  \int_{0}^{\infty}\bigg \{ \int_0^{\infty}  f(x,a,k) \bigg [\rho_{-k}(x,a,t^-))-\rho_k(x,a,t^-) \bigg ]da \bigg \}dx \nonumber \\
  &\quad +h(t)\int_0^{\infty} q(a,t)[f(0^-,a,1) -f(0^-,a,-1) ]da.
\label{oo}
\end{align}
Imposing continuity of the functions $f(x,a,k)$ across $x=0$ and using the fact that they are otherwise arbitrary gives the SPDE
\begin{subequations}
\begin{align}
  &\frac{\partial \rho_k}{\partial t}= -vk\frac{\partial}{\partial x} \rho_k(x,a,t)+h(t) [\rho_{-k}(x,a,t^-)-\rho_k(x,a,t^-)],\ k=\pm 1,\ x >0,\\
   &0=h(t)q(a,t)-v\rho_1(0^+,a,t)  ,\\
   &\frac{\partial q(a,t)}{\partial t}+\frac{\partial q(a,t)}{\partial a}=q(0,t)\delta(a)-h(t) q(a,t)+v\rho_{-1}(0^+,a,t).
 \label{SPDEstick}
\end{align}
\end{subequations}
Finally, taking expectations with respect to the Poisson process $N(t)$ and setting
\begin{equation}
\calP_k(x,a,t)=\E[\rho_k(x,a,t)],\quad \calQ(a,t)=\E[q(a,t)],
\end{equation}
we obtain the CK equation for the occupation time propagator:
\begin{subequations}
\begin{align}
\label{CKsticka2}
 &\frac{\partial \calP_k(x,a,t)}{\partial t}=-vk\frac{\partial \calP_k(x,a,t)}{\partial x}+\alpha [\calP_{-k}(x,a,t)-\calP_k(x,a,t)],\quad k=\pm 1,\ x \in \R,\\
\label{CKstickb2}
 &\alpha \calQ(a,t)=v\calP_1(0^+,a,t),\\
 & \frac{\partial \calQ(a,t)}{\partial t}+ \frac{\partial \calQ(a,t)}{\partial a}=\calQ(0,t)\delta(a)-\alpha \calQ(a,t)+v\calP_{-1}(0^+,a,t).
\label{CKstickc2}
\end{align}
\end{subequations}
Equations (\ref{CKsticka2})--(\ref{CKstickc2}) are precisely the equations posited in Ref. \cite{Bressloff23}. Here we have shown how the stochastic calculus of RTPs provides a rigorous mathematical framework for deriving these equations

Directly solving equations (\ref{CKsticka2})--(\ref{CKstickc2}) is a non-trivial problem. However, as highlighted in Ref. \cite{Bressloff23}, if we Laplace transform the equations with respect to the occupation time $a$, then we obtain the simpler boundary value problem (BVP)
\begin{subequations}
\begin{align}
\label{CKsticka2LT}
 &\frac{\partial \calPP_k(x,z,t)}{\partial t}=-vk\frac{\partial \calPP_k(x,z,t)}{\partial x}+\alpha [\calPP_{-k}(x,z,t)-\calPP_k(x,z,t)],\ k=\pm 1,\ x >0,\\
\label{CKstickb2LT}
 &\alpha \calQQ(z,t)=v\calPP_1(0^+,z,,t),\\
 & \frac{\partial \calQ(z,t)}{\partial t}=-(\alpha+z) \calQQ(z,t)+v\calPP_{-1}(0^+,z,t),
\label{CKstickc2LT}
\end{align}
\end{subequations}
 where
\begin{subequations}
\begin{align}
 \calPP(x,z,t)&=\int_0^{\infty} \e^{-za}  \calP(x,a,t) da,\ \calQQ(z,s )=\int_0^{\infty} \e^{-za} \calQQ(a,t) da.
\end{align}
\end{subequations}
Equations (\ref{CKsticka2LT})--(\ref{CKstickc2LT}) are equivalent to the CK equation for an RTP in $\R^+$ with a sticky boundary at $x=0$ and a constant rate of absorption $\kappa=z$. Given the solution to this BVP, we can then determine the solution to the propagator CK equation by inverting the Laplace transforms. Finally, substituting the result into the marginal density equations
(\ref{peep}) shows that
\begin{align}
\label{peep2}
   p^{\Psi}(x,t)&=\int_0^{\infty}\Psi(a)  {\rm LT}^{-1}[\calPP ](x,a,t) da  , \quad Q^{\Psi}(t)=\int_0^{\infty}\Psi(a) {\rm LT}^{-1}[\calQQ](a,t)da .
\end{align}
In Ref. \cite{Bressloff23} we solved the analog of equations (\ref{CKsticka2LT})--(\ref{CKstickc2LT}) for an RTP in the finite interval $[0,L]$ with sticky boundaries at both ends. We then calculated the MFPT for absorption as a function of the constant absorption rate $z$. Finally, we used equations (\ref{peep2}) to calculate the MFPT in the case of more general occupation time threshold distributions $\Psi(a)$.

\subsection{Sticky boundary at $x=0$ with resetting in the bulk domain.}
Another useful feature of the probabilistic formulation is that one can straightforwardly build more complicated models, at least when the various noise sources are independent. In order to illustrate this point, let us return to the example of a non-absorbing sticky boundary but now include the stochastic resetting protocol $(X(t),\sigma(t))\rightarrow (X_0,\sigma_0)$ at a rate $r$. Following section 3, we incorporate resetting into the SDE (\ref{dXsticky0}) by introducing a second Poisson process $\overline{N}(t)$ such that
\begin{subequations}
\begin{align}
\label{dXstickyr}
 dX(t)&=\bigg [v\sigma(t)dt+(x_0-X(t))d\overline{N}(t) \bigg ]{\bf 1}_{X(t)>0},\\
   d\sigma(t)&=  [\sigma_0-\sigma(t^-)]]{\bf 1}_{X(t)>0}\ d\overline{N}(t)-2  \sigma(t^-)dN(t)
  \end{align}
  \end{subequations}
  for all $ X(t)\in \Omega_{\epsilon}$. We are assuming that resetting only occurs in the bulk domain, that is, when the particle is stuck at the wall (lies within the boundary layer) it cannot reset. Performing the various steps outlined at the end of section 2, we obtain a modified version of equations (\ref{CKsticka})--(\ref{CKstickc}) that includes resetting:
\begin{subequations}
\begin{align}
 &\frac{\partial p_k(x,t)}{\partial t}=-vk\frac{\partial p_k(x,t)}{\partial x}+\alpha [p_{-k}(x,t)-p_k(x,t)]-rp_k(x,t)+r\delta(x-x_0)\delta_{k,\sigma_0}, \nonumber \\
 & \qquad  k=\pm 1,\quad x >0,\label{CKstickar}\\
 &\alpha Q(t)=vp_1(0^+,t),
 \label{CKstickbr}\\
 & \frac{dQ(t)}{dt}=-\alpha Q(t)+vp_{-1}(0^+,t),
\label{CKstickcr}
\end{align}
\end{subequations}

Modeling stochastic resetting in the case of a partially absorbing sticky boundary is more subtle. As previously highlighted for encounter-based models of diffusion with resetting \cite{Bressloff22c,Bressloff22d}, one has to specify whether or not the occupation time within a partially absorbing domain also resets. One of the motivations for considering encounter-based models is that the diffusing particle can alter the reactivity of an absorbing substrate. In this case one would not expect the occupation time to reset when the particle position resets. However, an alternative scenario involves the substrate modifying an internal state $\calU(t)$ of the particle that affects the probability of absorption. More specifically, $\calU(t)=F(A(t))$ with $F(0)=0$ and $F'(a)> 0$ for all $a\geq 0$. Hence, $\calU(t)$ is a strictly monotonically increasing function of the occupation time. The stopping condition for absorption is then taken to be of the form
\begin{equation}
\label{TU}
{\mathcal T}=\inf\{t>0:\ \calU(t) >\widehat{\calU}\},
\end{equation}
where $\widehat{\calU}$ is a random variable with probability density $p(u)$. Since $F(a)$ is strictly monotonic, we have the equivalent stopping condition
\begin{equation}
\label{TA2}
{\mathcal T}=\inf\{t>0:A(t) >\widehat{A}=F^{-1}(\widehat{\calU})\}.
\end{equation}
The one major difference from the substrate modification scenario is that one could now plausibly assume that $\calU(t)$ and thus $A(t)$ also reset:
$(X(t),A(t),\sigma(t))\rightarrow (X_0,0,\sigma_0)$. From a mathematical perspective, allowing the occupation time to reset significantly simplifies the analysis since resetting can be treated as a renewal process. The effects of stochastic resetting on an RTP with sticky boundary conditions will be explored further elsewhere.

\section{Stochastic entropy production for an RTP}

So far we have focused on using the stochastic calculus of jump processes and generalised It\^o lemmas to derive the CK equation for an RTP with resetting or a partially absorbing sticky boundary. In this section we turn to another application of the probabilistic approach, namely, stochastic entropy production. In recent years there has been considerable interest in stochastic thermodynamics, which extends familiar thermodynamical quantities such as heat, work and entropy to non-equilibrium mesoscopic systems  \cite{Seifert05,Sekimoto10,Seifert12,Cocconi20,Peliti21,Roldan23}.  Evaluating entropy production along individual trajectories of the underlying stochastic process can result in both increases and decreases in the stochastically-defined entropy. However, the second law of thermodynamics is recovered after taking appropriate averages with respect to the ensemble of stochastic trajectories. The resulting average rate of entropy production then
quantifies the degree of departure from thermodynamic equilibrium. This is particularly important for
non-equilibrium processes that are characterised by a positive average entropy production at steady-state,
which is  the rate at which heat is dissipated into the environment.  

 Most work on the thermodynamics of continuous stochastic processes have focused on Brownian motion. There have been a few recent studies of entropy production for RTPs \cite{Cocconi20,Frydel22,Angelani24}, but these analyse the Gibbs-Shannon entropy, which for an RTP in $ \R$ is defined according to
\begin{equation}
S_{\rm GS}(t):=-\sum_{k=\pm 1} \int_{\R} p_k(x,t)\ln p_k(x,t)dx,
\label{GS}
\end{equation}
The corresponding average rate of entropy production is obtained by differentiating both sides of equation (\ref{GS}) with respect to $t$, imposing the condition that $p_k(x,t)$ is a solution of the CK equation (\ref{CKH}), and then separating out the contribution from the environmental entropy. One advantage of working with the full stochastic entropy along an individual trajectory is that it maintains information about fluctuations that is lost after averaging \cite{Peliti21}. Moreover, as we will show below, the expression for the average rate of entropy production is recovered by taking double expectations with respect to the white noise process and the discrete switching process. One limitation regarding both definitions of entropy is that obtaining explicit solutions for the individual components $p_k(x,t)$ of the underlying CK equation (\ref{CK}) is non-trivial.

\subsection{RTP with diffusion in $\R$.}

Let us define the instantaneous system entropy by
\begin{equation}
S_{\rm sys}(t)=-\ln p_{\sigma(t)}(X(t),t),
\label{Ssys}
\end{equation}
with $p_k(x,t)$ evolving according to equation (\ref{CK}). Taking differentials of both sides with respect to $t$ using the generalised It\^o formula (\ref{lemma0}) gives
\begin{align}
  dS_{\rm sys}(t)  &= -\frac{1}{ p_{\sigma(t)}(X(t),t)}\frac{\partial  p_{\sigma(t)}(X(t),t)}{\partial t}dt -d_x[\ln p_{\sigma(t)}(X(t),t)]\nonumber \\
  &\quad 
-\bigg [\ln p_{-\sigma(t^-)}(X(t),t^-)-\ln p_{\sigma(t^-)}(X(t),t^-)
\bigg ]dN(t) ,
\end{align}
with $N(t)$ the Poisson process generating direction reversals and
\begin{align}
\label{org0}
  &-d_x[\ln p_{\sigma(t)}(X(t),t)]=-\frac{v\sigma(t) dt+  \sqrt{2D}dW(t)}{p_{\sigma(t)}(X(t),t)} \frac{\partial  p_{
\sigma(t)}(X(t),t)}{\partial x} \\
  & -D\bigg \{\frac{1}{p_{\sigma(t)}(X(t,t)}  \frac{\partial ^2 p_{\sigma(t)}(X(t),t)}{\partial x^2}-\left [\frac{1}{p_{\sigma(t)}(X(t,t)} \frac{\partial  p_{
\sigma(t)}(X(t),t)}{\partial x}\right ]^2\bigg \}dt.\nonumber 
 \end{align}
 Introducing the probability currents (\ref{Jk})
 and noting that 
 \[\frac{J_k^2}{Dp_k^2}=D\left [\frac{1}{p_k}\frac{\partial p_k}{\partial x}\right ]^2-\frac{2vk}{p_k}\frac{\partial p_k}{\partial x}+\frac{v^2}{D},\]
 we find that
 \begin{align}
  dS_{\rm sys}(t)  &= -\frac{1}{ p_{\sigma(t)}(X(t),t)}\frac{\partial  p_{\sigma(t)}(X(t),t)}{\partial t}dt -\frac{v\sigma(t)J_{\sigma(t)}(X(t),t)}{Dp_{\sigma(t)}(X(t),t)}dt +\frac{J^2_{\sigma(t)}(X(t),t)}{Dp^2_{\sigma(t)}(X(t),t)}dt \nonumber \\
  & \quad - \frac{D}{p_{\sigma(t)}(X(t,t)}  \frac{\partial ^2 p_{\sigma(t)}(X(t),t)}{\partial x^2}dt- \frac{\sqrt{2D}dW(t)}{p_{\sigma(t)}(X(t),t)} \frac{\partial  p_{
\sigma(t)}(X(t),t)}{\partial x} \nonumber \\
  &\quad 
-\bigg [\ln p_{-\sigma(t^-)}(X(t),t^-)-\ln p_{\sigma(t^-)}(X(t),t^-)
\bigg ]dN(t) .\label{dSsys} 
\end{align}

The average rate of system entropy production can be calculated by taking a double expectation of each term in equation (\ref{dSsys}) with respect to the white noise and direction reversal processes. 
This gives
\begin{align}
& \calR_{\rm sys}(t)\nonumber \\
&\quad :=\left \langle \E\left [\frac{dS_{\rm sys}(t)}{dt}\right ]\right \rangle  =\sum_{k=\pm 1}\int_{\R}p_k(x,t)\bigg\{  -\frac{1}{ p_{k}(x,t)}\frac{\partial  p_{k}(x,t)}{\partial t}-\frac{kvJ_{k}(x,t)}{ D p_{k}(x,t)} +\frac{J_{k}^2(x,t)}{D  p_{k}^2(x,t)} \nonumber \\
 &\quad - \frac{D}{ p_k(x,t)}\frac{\partial^2  p_{k}(x,t)}{\partial x^2} \bigg \}dx +{\mathcal I}(t)\nonumber \\
 &=-\sum_{k=\pm 1}\int_{\R} \frac{kvJ_{k}(x,t)}{ D }dx+\sum_{k=\pm 1} \int_{\R}  \frac{J_k^2(x,t)}{D p_k(x,t)}dx +{\mathcal I}(t),
\label{Rsys}
\end{align}
where
\begin{align}
{\mathcal I}(t)&:= \left \langle \E\left [ \sum_{n \geq 1}\delta(t-T_{n})\ln\bigg ( p_{\sigma(t^-)}(X(t),t)/p_{-\sigma(t^-)}(X(t),t)\bigg )\right ]\right \rangle \nonumber \\
& =\alpha \sum_{k=\pm 1} \int_{\R}  p_k(x,t) \ln [p_{k}(x,t^-)/p_{-k}(x,t^-)]dx\nonumber \\
&  =\alpha \int_{\R} [p_1(x,t)-p_{-1}(x,t)]\ln[p_1(x,t)/p_{-1}(x,t)]dx.
\label{Roo}
\end{align}
There is also exchange of heat energy with the environment, which results in the production of environmental entropy at a rate
\begin{align}
 \calR_{\rm env}(t)  =\sum_{k=\pm 1}\int_{\R} \frac{kvJ_{k}(x,t)}{ D }dx=\frac{v}{D}\int_{\R} (J_1(x,t)-J_{-1}(x,t))dx.
\label{Renv}
\end{align}
Combining equations (\ref{Rsys}), (\ref{Roo}) and (\ref{Renv}) shows that
\begin{align}
 \calR_{\rm sys}(t)  &= \calR_{\rm tot}(t) - \calR_{\rm env}(t) 
\label{Rsys2}
\end{align}
where $ \calR_{\rm tot}(t) $ is the total rate of entropy production:
\begin{align}
 \calR_{\rm tot}(t)&=\sum_{k=\pm 1} \int_{\R}  \frac{J_k^2(x,t)}{D p_k(x,t)}dx +\alpha \int_{\R} [p_1(x,t)-p_{-1}(x,t)]\ln[p_1(x,t)/p_{-1}(x,t)]dx.
\label{Rtot}
\end{align}
Since $(a-b)$ and $\ln (a/b)$ have the same sign, it follows that  $\calR_{\rm tot}(t)\geq 0$, which is a version of the second law of thermodynamics. 

Comparison with the results of Ref. \cite{Angelani24}) establishes that
\begin{equation}
\left \langle \E\left [\frac{dS_{\rm sys}(t)}{dt}\right ]\right \rangle=\frac{d}{dt}S_{\rm GS}(t).
\end{equation}
Moreover, the steady-state total entropy production rate can be obtained without explicitly solving the steady-state CK equation.
This follows from the observation that $\calR_{\rm sys}^*=0$ and, hence,
\begin{equation}
\calR_{\rm tot}^*=\calR_{\rm env}^*=\frac{v}{D}\int_{\R} (J_1^*(x)-J_{-1}^*(x))dx,
\end{equation}
where $^*$ denotes a steady-state solution. From the definition of $J_k$ in equation (\ref{Jk}) one finds that \cite{Angelani24}
\begin{equation}
 \calR_{\rm tot}^* =\frac{v}{D}\int_{\R} \bigg (v[p_1^*(x)+p_{-1}^*(x)]-D\partial_x[p_1^*(x)-p_{-1}^*(x)]\bigg )dx=\frac{v^2}{D}.
\end{equation}
This result  immediately follows from the normalisation condition $\int_{\R}[p_1^*(x)+p_{-1}^*(x)]dx=1$ and the vanishing of the solution at $\pm \infty$. In principle, we could determine fluctuations in the steady-state entropy production $\calR_{\rm tot}^*$ by setting $p_{\sigma(t)}(X(t),t) =p^*_{\sigma(t)}(X(t))$ in equation (\ref{Ssys}) and numerically evaluating how the entropy changes along a sample trajectory. However, this requires an explicit solution for $p_k^*(x)$.

\subsection{RTP in $\R$ with stochastic resetting and $D=0$.}

One example where the non-equilibrium stationary state (NESS) $p_k^*(x)$ can be calculated explicitly is an RTP in $\R$ with stochastic resetting and no diffusion \cite{Evans18}. The corresponding CK equation is given by (\ref{CK0r}). Analogous to the previous example, we can determine changes in the stochastic entropy (\ref{Ssys}) along a sample trajectory using the generalised It\^o's lemma (\ref{lemmar}) with $D=0$. This gives
\begin{align}
  dS_{\rm sys}(t)  &= -\frac{1}{ p_{\sigma(t)}(X(t),t)}\frac{\partial  p_{\sigma(t)}(X(t),t)}{\partial t}dt -\frac{v\sigma(t)}{p_{\sigma(t)}(X(t),t)} \frac{\partial  p_{
\sigma(t)}(X(t),t)}{\partial x}dt \nonumber \\
  &\quad 
-\bigg [\ln p_{\sigma_0}(x_0,t)-\ln p_{\sigma(t^-)}(X(t^-),t^-)\bigg ]d\overline{N}(t)\nonumber \\
 &\quad -\bigg [\ln p_{-\sigma(t^-)}(X(t),t^-)-\ln p_{\sigma(t^-)}(X(t),t^-)
\bigg ]dN(t) ,\label{dSsysr} 
\end{align}
with $N(t)$ and $\overline{N}(t)$ the Poisson process generating direction reversals and resetting, respectively. Since the switching rates between the two velocity states are the same, there is no change in free energy associated with the transitions (assuming detailed balance holds). This implies that there is no dissipation of heat into the environment from switching. We also assume that no heat is dissipated when the particle resets its position and velocity state.\footnote{It is important to note that instantaneous resetting is a mathematical idealisation of a real physical process so determining its thermodynamics from first principles is not possible.} Hence, $S_{\rm sys}(t) $ is also the total stochastic entropy.

In Ref. \cite{Evans18} the NESS for the marginal probability density $p^*(x)=p_{1}^*(x)+p_{-1}^*(x)$ was calculated in the case of the symmetric resetting rule $\sigma(t)\rightarrow \pm 1$ with equal probability. In the appendix we extend this calculation to determine the individual components $p_{\pm}^*(x)$ for the asymmetric resetting condition $\sigma(t)\rightarrow \sigma_0$. We summarise the results here. If $\sigma_0=1$ we have
\begin{align}
\left .\p_{1}^*(x)\right |_{\sigma_0=1}&=\frac{r}{v}\e^{-\lambda_r(x-x_0)}\Theta(x-x_0)+\frac{
r}{2\Lambda_r}\frac{\alpha^2}{v^3}\left [\frac{\e^{-\Lambda_r (x-x_0)}}{\lambda_r-\Lambda_r}- \frac{2\Lambda_r\e^{-\lambda_r (x-x_0)}}{\lambda_r^2-\Lambda_r^2}\right ]\Theta(x-x_0)\nonumber \\
&\quad +\frac{r}{2\Lambda_r}\frac{\alpha^2}{v^3}\frac{\e^{-\Lambda_r |x-x_0|}}{\lambda_r+\Lambda_r}\Theta(x_0-x).
\label{1pssa}
\end{align}
and
\begin{equation}
\label{1pssb}
\left .\p_{-1}^*(x)\right |_{\sigma_0=+1}=\frac{\alpha}{v^2}\frac{r}{2\Lambda_r}\e^{-\Lambda_r|x-x_0|}.
\end{equation}
Similarly, if $\sigma_0=-1$ then
\begin{equation}
\label{-1pssa}
\left .\p_{1}^*(x)\right |_{
\sigma_0=-1}=\frac{\alpha}{v^2}\frac{r}{2\Lambda_r}\e^{-\Lambda_r|x-x_0|}.
\end{equation}
and
 \begin{align}
\left .\p_{-1}^*(x)\right |_{\sigma_0=-1}&=\frac{r}{v}\e^{-\lambda_r|x-x_0|}\Theta(x_0-x)
 +\frac{r}{2\Lambda_r}\frac{\alpha^2}{v^3}\left [\frac{\e^{-\Lambda_r |x-x_0|}}{\lambda_r-\Lambda_r}-\frac{2\Lambda_r\e^{-\lambda_r |x-x_0|}}{\lambda_r^2-\Lambda_r^2}\right ]\Theta(x_0-x)\nonumber \\
&\quad +\frac{r}{2\Lambda_r}\frac{\alpha^2}{v^3}\frac{\e^{-\Lambda_r |x-x_0|}}{\lambda_r+\Lambda_r}\Theta(x-x_0).
\label{-1pssb}
\end{align}
Here
\begin{equation}
\Lambda_r =\frac{\sqrt{r(r+2\alpha)}}{v},\quad \lambda_r= \frac{\alpha+r}{v}.
\end{equation}

 \begin{figure}[t!]
\centering
\includegraphics[width=14cm]{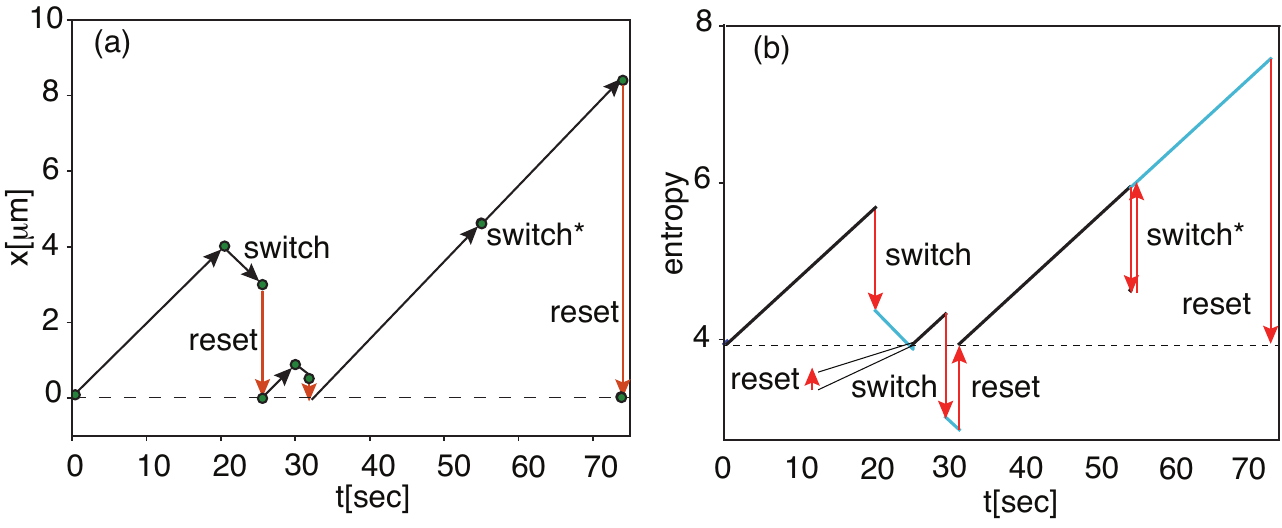}
\caption{(a) Sample trajectories of the position $X(t)$ of an RTP with resetting. Here $x_0=0$ and $\sigma_0=1$. The switching rate and resetting rate are both taken to be 0.25 s$^{-1}$ and $v=1\mu$m s$^{-1}$. The notation ``switch*'' indicates two direction reversals that occur almost simultaneously. (b) Corresponding trajectory of the stochastic entropy $S(X(t),\sigma(t)$.}\label{fig3}
\end{figure}

\begin{figure}[b!]
\centering
\includegraphics[width=14cm]{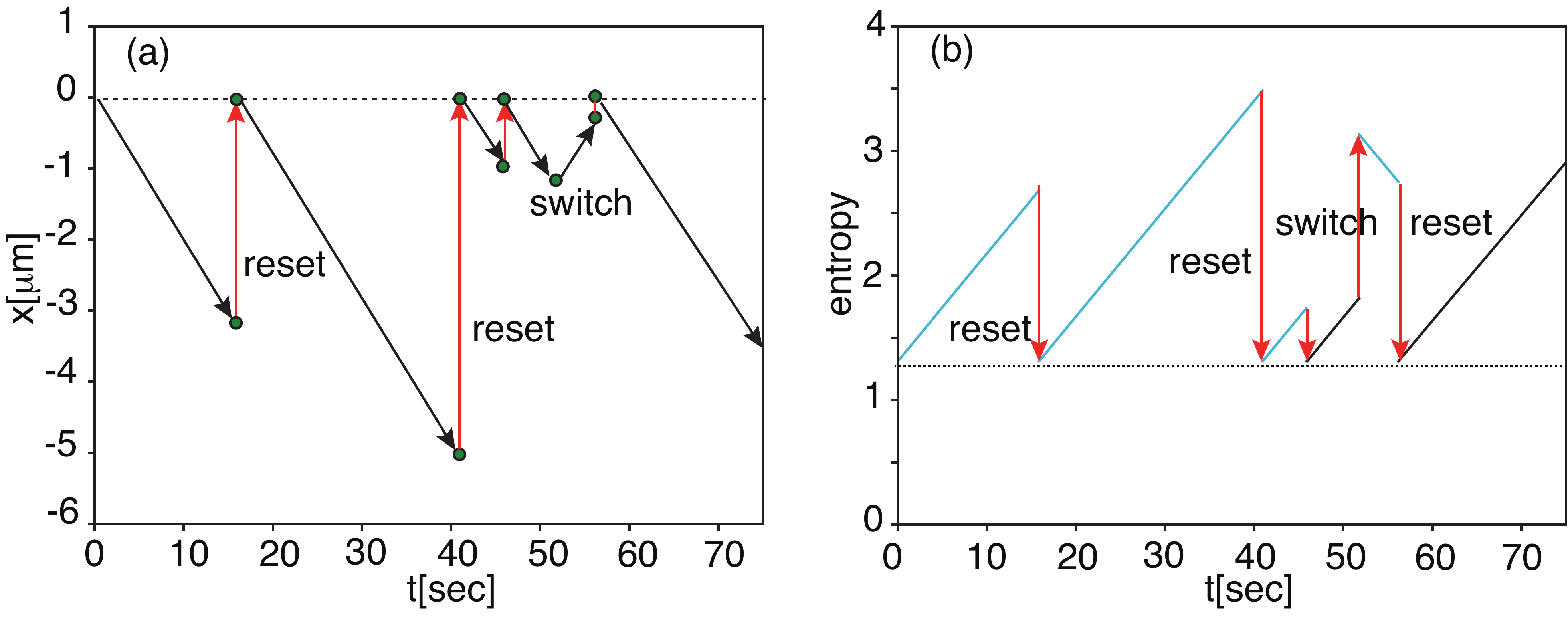}
\caption{Same as Fig. \ref{fig3} except that $\sigma_0=-1$.}\label{fig4}
\end{figure}

Given the above NESS, we set $p_k=p_k^*(x)$ in equation (\ref{dSsysr}) such that
\begin{align}
  dS_{\rm sys}^*(t)  &= -\frac{v\sigma(t)}{p^*_{\sigma(t)}(X(t))} \frac{\partial  p^*_{
\sigma(t)}(X(t))}{\partial x}dt  -\bigg [\ln p^*_{\sigma_0}(x_0)-\ln p^*_{\sigma(t^-)}(X(t^-))\bigg ]d\overline{N}(t)\nonumber \\
 &\quad -\bigg [\ln p^*_{-\sigma(t^-)}(X(t)) -\ln p^*_{\sigma(t^-)}(X(t))
\bigg ]dN(t) .\label{dSstar} 
\end{align}
The terms multiplying the Poisson differentials $d\overline{N}(t)$ and $dN(t)$ represent, respectively, the jumps in entropy at the resetting times $\overline{\calT}_n$ and reversal times $\calT_n$, $n\geq 1$. The first term on the right-hand side determines the continuous change in entropy due to the ballistic motion between jumps. Suppose that $\calS_m\leq t<\calS_{m+1}$ where $\calS_m$ is the time of the $m$th jump event. Let $X_m=X(\calS_m)$ and $\sigma_m=\sigma(\calS_m)$. It follows that $X(t)=X_m+v(t-\calS_m)$ for all $t\in [\calS_m,\calS_{m+1})$, and we can write
\begin{align}
\frac{dS}{dt}=- v\sigma_m \frac{\partial  \ln p^*_{
\sigma_m}(X_m+v(t-\calS_m))}{\partial x}=-  \frac{\partial  \ln p^*_{
\sigma_m}(X_m+v(t-\calS_m))}{\partial t}
\end{align}
for $\calS_m\leq t < \calS_{m+1}$. Integrating both sides with respect to $t$ shows that the change in entropy between jumps is
\begin{align}
\Delta S_m= -\ln \bigg [\frac{p^*_{\sigma_m}(X_m+v(\calS_{m+1}-\calS_m))}{p^*_{\sigma_m}(X_m)}\bigg ].
\end{align}

Figs. \ref{fig3}(a,b) show sample trajectories of the position $X(t)$ and entropy $S(t)=S(X(t),\sigma(t)$, respectively, with $x_0=0$ and $\sigma_0=1$. Sample trajectories for $\sigma_0=-1$ are shown in Figs. \ref{fig4}(a,b). It can be seen that each switching event leads to a reversal of direction $\sigma(t)\rightarrow -\sigma(t)$, whereas each resetting event leads to the instantaneous transition $(X(t),\sigma(t))\rightarrow (0,1)$. The corresponding entropy jumps at both types of event. with $S(t)\rightarrow S(0,1)$ in the particular case of resetting. Note that the trajectories of the entropy between jumps appear to be straight lines. However, plotting $p_{\sigma(t)}(X(t)$ rather than its logarithm shows that they are actually curved.

\paragraph{\bf Average rate of entropy production.}The average rate of system entropy production can be calculated by taking expectations of each term in equation (\ref{dSsysr}) with respect to the white noise and Poisson processes. 
This gives
\begin{align}
\label{Rsysr}
 \calR_{\rm sys}(t)&:=\left \langle \E\left [\frac{dS_{\rm sys}(t)}{dt}\right ]\right \rangle  \\
 &=\sum_{k=\pm 1}\int_{\R}p_k(x,t)\bigg\{  -\frac{1}{ p_{k}(x,t)}\frac{\partial  p_{k}(x,t)}{\partial t}-\frac{kv}{  p_{k}(x,t)} \frac{\partial  p_{k}(x,t)}{\partial x}\bigg \} \nonumber \\
 &\quad +\alpha \int_{\R} [p_1(x,t)-p_{-1}(x,t)]\ln[p_1(x,t)/p_{-1}(x,t)]dx \nonumber \\
 &\quad +r \int_{\R} \bigg \{p_1(x,t)\ln\left [\frac{p_1(x,t)}{p_{\sigma_0}(x_0,t)}\right ]+p_{-1}(x,t)\ln \left [\frac{p_{-1}(x,t)}{p_{\sigma_0}(x_0,t)}\right ]\bigg \}dx.\nonumber
\end{align}
The first two terms on the right-hand side vanish identically. Since the switching rates between the two velocity states are the same, there is no change in free energy associated with the transitions (assuming detailed balance holds). This implies that there is no dissipation of heat into the environment from switching. Following previous studies of Brownian motion with resetting \cite{Fuchs16,Mori23}, we identify the final term as a resetting entropy production rate $\calR_{\rm res}$ and express the second law as
\begin{equation}
 \calR_{\rm sys}(t)-\calR_{\rm sys}(t)=\alpha \int_{\R} [p_1(x,t)-p_{-1}(x,t)]\ln[p_1(x,t)/p_{-1}(x,t)]dx\geq 0.
\end{equation}

\setcounter{equation}{0}
\section{Population of RTPs with global resetting}
Our final application of the probabilistic formulation concerns deriving a global density equation for a population of RTPs, which we develop along analogous lines to recent work on Brownian gases \cite{Bressloff24a,Bressloff24b}. Let $X_j(t)\in \R$ and $\sigma_j(t)\in \{1,-1\}$ denote the position and velocity state of the $j$th RTP at time $t$, $j=1,\ldots,\calM$, where $\calM$ is the number of particles. Suppose that 
each RTP resets to its initial state $(x_{j,0},\sigma_{j,0})$ at a fixed Poisson rate $r$. For the sake of illustration, consider a global update rule where all of the particles simultaneously reset at the sequence of times $\overline{\calT}_n$ generated from a common Poisson process $\overline{N}(t)$ \cite{Nagar23}. On the other hand, the $j$th particle switches its velocity at the sequence of times ${\calT}_{j,n}$ generated from an independent Poisson process $N_j(t)$, $j=1,\ldots,\calM$.
The population versions of equations (\ref{dXr}) and (\ref{dsigr}) are
\begin{equation}
dX_j(t)=v\sigma_j(t)dt+\sqrt{2D}dW_j(t)+(x_{j,0}-X_j(t^-))d\overline{N}(t),
\label{dXpop}
\end{equation}
and
\begin{align}
d\sigma_j(t)&=  [\sigma_{j,0}-\sigma_j(t^-)]d\overline{N}(t)-2  \sigma_j(t^-)dN_j(t).
\label{dsigpop}
\end{align}
where ${W}_j(t)$ are independent Wiener processes. One can then derive a multi-particle analog of the SPDE (\ref{SPDEH}) for the population-based empirical measures
\begin{eqnarray}
\rho_k(x,t)=\frac{1}{\calM}\sum_{j=1}^{\calM}\delta(x-X_j(t))\delta_{k,\sigma_j(t)},\quad k=\pm 1,
\end{eqnarray} 
by following the steps outlined at the end of section 2. First, we introduce an arbitrary smooth test function $f(x,k)$ of compact support and write
\begin{align}
&\sum_{k=\pm 1}\left [ \int_{\R}  f(x,k)\frac{\partial \rho_k(x,t)}{\partial t}dx \right ] 
=\frac{1}{\calM}\sum_{j=1}^{\calM}\frac{df(X_j(t),\sigma_j(t)) }{dt} .  
\end{align}
Evaluating the right-hand side using the generalised It\^o's lemma (\ref{lemmar}), 
integrating by parts the various terms involving derivatives of $f$ and using the fact that $f$ is arbitrary yields the following SPDE for $\rho_k$:
\begin{align}
 \label{rholoc}
 \frac{\partial \rho_k(x,t)}{\partial t} 
 &=-\frac{2\sqrt{D}}{\calM}\sum_{j=1}^{\calM}  \bigg [{ \xi}_j(t) \frac{\partial \delta(x-X_j(t))}{\partial x} \bigg ]+D\frac{\partial^2 \rho_k(x,t)}{\partial x^2}-vk\frac{\partial \rho_k(x,t)}{\partial x} \\
 &\quad  +  \frac{1}{\calM}\sum_{j=1}^{\calM}h_j(t)\delta(x-X_j(t))  \bigg [ \delta_{k,-\sigma_j(t^-)}-\delta_{k,\sigma_j(t^-)}\bigg ]+\overline{h}(t)[\rho_{k,0}(x)-\rho_k(x,t^-)]  . \nonumber
\end{align}
where $d{W}_j(t)={\xi}_j(t)dt$,
 \begin{eqnarray}
\label{hj}
dN_j(\tau):=h_j(\tau)d\tau  =\sum_{n=1}^{\infty}\delta(\tau-{\calT}_{j,n})d\tau ,\qquad  \rho_{k,0}(x)=\frac{1}{\calM}\sum_{j=1}^{\calM}\delta(x-x_{j,0})\delta_{k,\sigma_{j,0}}.
\end{eqnarray}

As it stands, (\ref{rholoc}) is not a closed system of equations for the pair of densities $\rho_k(x,t)$.
However, averaging with respect to the set of independent white noise processes $\xi_j(t)$ and   Poisson processes $N_j(t)$, $j=1,\ldots,\calM$, yields a closed stochastic equation for 
$\Phi_k(\x,t)=\E_N[\langle \rho_k(\x,t)\rangle]$. (Here $\E_N[\cdot]$ indicates that we average with respect to the $N_j(t)$, $j=1,\ldots,\calM$, but not $\overline{N}(t)$.) This follows from noting that the first term on the right-hand side of Eq. (\ref{rholoc})  has zero mean and
\begin{eqnarray}
&& \frac{1}{\calM}\E_N\bigg [\sum_{j=1}^{\calM} \left \langle \delta(x-X_j(t))  \bigg ( \delta_{k,-\sigma_j(t^-)}-\delta_{k,\sigma_j(t^-)}\bigg )h_j(t)\right \rangle \bigg ] \nonumber\\
&&\quad =
\frac{1}{\calM}\E_N\bigg [ \sum_{j=1}^{\calM}\left \langle \delta(x-X_j(t))  \bigg ( \delta_{k,-\sigma_j(t^-)}-\delta_{k,\sigma_j(t^-)}\bigg )\right \rangle \bigg ]\E_N[h_j(t)]\nonumber \\
&&\quad = \alpha \bigg (\E_N[\langle \rho_{-k}(\x,t)\rangle]-\E_N[\langle \rho_k(\x,t)\rangle]\bigg )=\alpha[\Phi_{-k}(x,t)-\Phi_k(x,t)].
\end{eqnarray}
Hence, $\Phi_k(x,t)$ satisfies the stochastic CK equation
\begin{align}
 \frac{\partial \Phi_k(x,t)}{\partial t}&=-vk\frac{\partial \Phi_k(x,t)}{\partial x}+D\frac{\partial^2 \Phi_k(x,t)}{\partial x^2}    +\alpha [\Phi_{-k}(x,t)-\Phi_k(x,t)] \nonumber \\
 &\quad +\overline{h}(t)\left [\rho_{k,0}(x) -\Phi_k(x,t^-)\right ],\quad k=\pm 1 .
\label{CKphi}
\end{align}
Equation (\ref{CKphi}) is a PDE with the stochastic resetting rule $\Phi_k(x,t^-)\rightarrow \rho_{k,0}(x)$ at the Poisson generated sequence of times $\overline{\calT}_{n}$. 

At first sight it would also make sense to average equation (\ref{CKphi}) with respect to the resetting process, which would yield a deterministic equation for $p_k(x,t)=\E_{\overline{N}}[\Phi_k(x,t)]$ where $\E_{\overline{N}}[\cdot ]$ denotes averaging with respect to $\overline{N}(t)$:
\begin{align}
 \frac{\partial p_k(x,t)}{\partial t}&=-vk\frac{\partial p_k(x,t)}{\partial x}+D\frac{\partial^2 p_k(x,t)}{\partial x^2}
 +\alpha [p_{-k}(x,t)-p_k(x,t)] \nonumber \\
 &\quad +r \left [ \rho_{k,0} -p_k(x,t) \right ] .
\label{CKphi2}
\end{align}
However, there is major difference between $\overline{N}(t)$ and the set of white noise processes $\xi_j(t)$ and Poisson processes $N_j(t)$. In the thermodynamic limit $\calM\rightarrow \infty$ we can use a version of the law of large numbers to deduce that fluctuations with respect to the local noise sources vanish. Such an argument does not hold for the global resetting process. One non-trivial consequence of global resetting is that it induces statistical correlations between the particles even though there are no particle-particle interactions. This follows from the observation that the product $C_{jk}(x,y,t)=\Phi_j(x,t)\Phi_k(y,t)$ satisfies the SPDE
\begin{align}
 \frac{\partial C_{jk}(x,y,t)}{\partial t}&=-vj\frac{\partial C_{jk}(x,y,t)}{\partial x}-vk\frac{\partial C_{jk}(x,y,t)}{\partial y}+D\frac{\partial^2 C_{jk}(x,y,t)}{\partial x^2}+D\frac{\partial^2 C_{jk}(x,y,t)}{\partial y^2}\nonumber \\
 &\quad +\alpha [C_{-j,k}(x,y,t)-C_{jk}(x,y,t)] +\alpha [C_{j,-k}(x,y,t)-C_{jk}(x,y,t)] \nonumber \\
 &\quad +\overline{h}(t)[\rho_{j,0}(x)\rho_{k,0}(y)-C_{jk}(x,y,t^-)] ,\quad k=\pm 1,
\label{CCK}
\end{align}
Averaging equations (\ref{CKphi}) and (\ref{CCK}) with respect to the global Poisson process $\overline{N}(t)$ and setting $\E[\Phi_j(x,t)\Phi_k(y,t)]=c_{jk}(x,y,t)$, we see that
\begin{align}
&\frac{\partial}{\partial t}\left [ c_{jk}(x,y,t)-p_j(x,t)p_k(y,t)\right ]\nonumber \\
&\quad =r[\rho_{j,0}(x)\rho_{k,0}(y)-c_{jk}(x,y,t)] +rp_j(x,t)[\rho_{k,0}(y)-p_{k}(y,t)]\nonumber \\
&\quad +rp_k(y,t)[\rho_{j,0}(x)-p_{j}(x,t)].
\end{align}
The above equation does not have the solution $ c_{jk}(x,y,t)=p_j(x,t)p_k(y,t)$ and hence,
\begin{equation}
\E[\Phi_j(x,t)\Phi_k(y,t)]\neq \E[\Phi_j(x,t)]\E[\Phi_k(y,t)].
\end{equation}

\begin{figure}[t!]
\centering
\includegraphics[width=14cm]{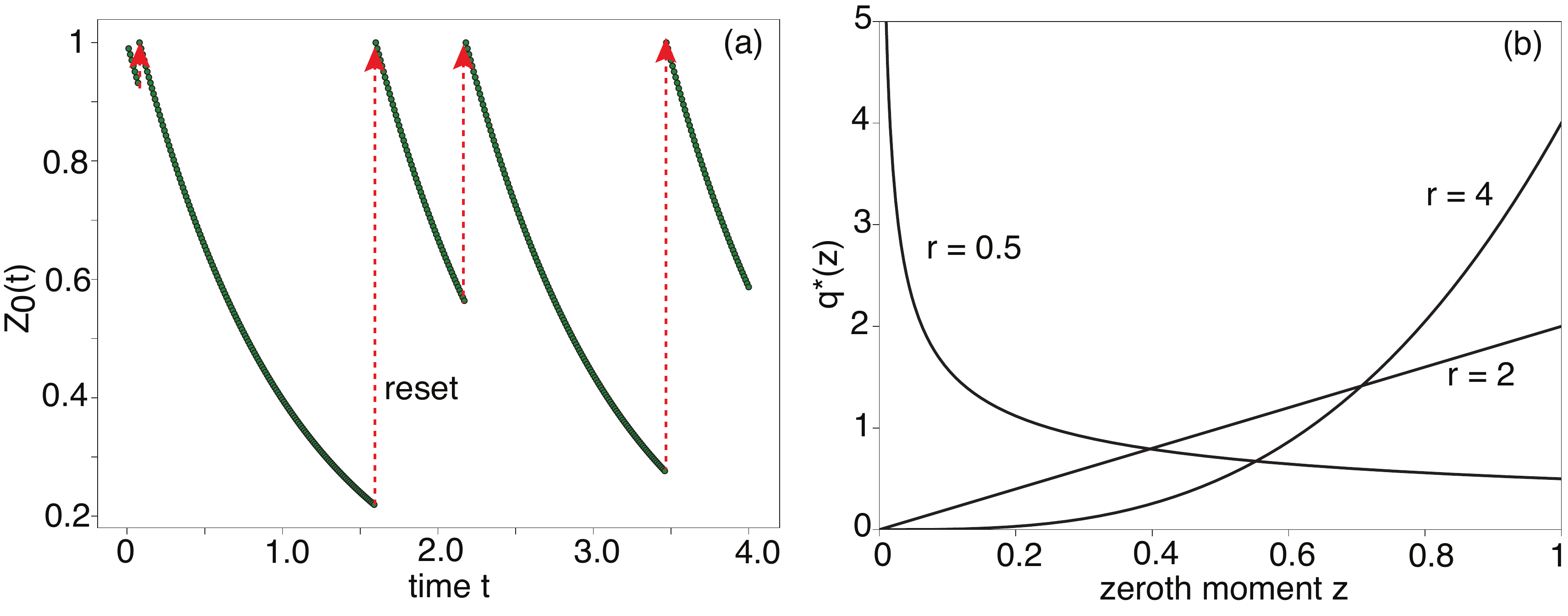}
\caption{(a) Sample plot of the zeroth moment $Z_0(t)=M_1^{(0)}(t)-M_{-1}^{(0)}(t)$ satisfying equation (\ref{Z0}) with $2\alpha=1$, $r=1$ and $\lambda_-^{(0)}=1$. (b) Corresponding NESS $q^*(z)$ given by equation (\ref{q0}) with $2\alpha=1$ and $\lambda_-^{(0)}=1$.}
\label{fig5}
\end{figure}

We conclude by highlighting an interesting mathematical structure contained within the stochastic CK equation (\ref{CKphi}), namely, the corresponding moment equations form a hierarchy of ODEs with resetting in which lower-order moments are embedded into the equations at higher orders. Define the $\ell$th order moment by
\begin{equation}
M^{(\ell)}_k(t)=\int_{\R}x^{\ell}\Phi_k(x,t)dx,\quad \ell \geq 0..
\end{equation}
and set
\begin{equation}
\lambda^{(\ell)}_k=\int_{\R}x^{\ell}\rho_{k,0}(x)dx=\frac{1}{\calM}\sum_{j=1}^{\calM} x_{j,0}^{\ell}\delta_{k,\sigma_{j,o}}.
\end{equation}
Note that $\lambda_+^{(0)}\equiv \lambda_1^{(0)}+\lambda_{-1}^{(0)}=1$.
Multiplying equation (\ref{CKphi}) by $x^{\ell}$ and integrating with respect to $x\in \R$ yields a hierarchy of equations for $k=\pm 1$:
\begin{subequations}
\label{hier}
\begin{align}
\frac{dM^{(\ell)}_k(t)}{dt}&= \ell vkM^{(\ell-1)}_k(t)+\ell(\ell-1)DM^{(\ell-2)}_k(t)
  +\alpha [M^{(\ell)}_{-k}(t)-M^{(\ell)}_k(t)] \nonumber \\
&\quad  +\overline{h}(t)[\lambda^{(\ell)}_k-M^{(\ell)}_k(t^-)],\quad \ell \geq 1,\\
\frac{dM^{(0)}_k(t)}{dt}&=\alpha [M^{(0)}_{-k}(t)-M^{0)}_k(t)]  +\overline{h}(t)[\lambda^{(0)}_k-M^{(0)}_k(t^-)]
\end{align}
\end{subequations}
The zeroth-order equation is relatively straightforward to analyze. Using the fact that $W_0(t)\equiv M_1^{(0)}(t)+ M_{-1}^{(0)}(t)=1$ by conservation of probability, equation (\ref{hier}b) reduces to a single equation for the difference $Z_0(t):= M_1^{(0)}(t)-M_{-1}^{(0)}(t)$:
\begin{align}
\frac{dZ_0(t)}{dt}&=   -2\alpha Z_0(t) -\overline{h}(t)[ \lambda_-^{(0)}-Z_0(t^-) ],
\label{Z0}
\end{align}
where $\lambda^{(0)}_{- }=\lambda^{(0)}_1- \lambda^{(0)}_{-1}$. For the sake of illustration, suppose that $\lambda_-^{(0)}>0$.
 The solution to equation (\ref{Z0}) then alternates between approaching $0$ from above and resetting to $ \lambda_-^{(0)}$, see Fig. \ref{fig5}(a). Hence, the support of $Z_0(t)$ is $ [0,\lambda_-^{(0)}]$. In addition, 
The CK equation corresponding to this ODE with resetting is
 \begin{align}
  \frac{\partial q(z,t)}{\partial t}&= 2\alpha \frac{\partial zq(z,t)}{\partial z} -rq(z,t) +r\delta(\lambda_-^{(0)}-z). \label{CKq}
\end{align}
with $q(z,t)dz=\P[z<Z_0(t)<z+dz]$.
The NESS $q^*(z)=\lim_{t\rightarrow \infty} q(z,t)$ satisfies the time-independent equation
\begin{align}
 -2\alpha z\frac{d q^*(z)}{d z}+(r-2\alpha)q^*(z)=r\delta(z-\lambda_-^{(0)})  , \label{CKq2}
\end{align}
which has the solution
\begin{equation}
\label{q0}
q^*(z)=\frac{r}{2\alpha \lambda_-^{(0)}}\left (\frac{z}{\lambda_-^{(0)}}\right )^{r/2\alpha-1}\Theta(\lambda_-^{(0)}-z).
\end{equation}
It can be checked that $\int_0^{\lambda_-^{(0)}}q^*(z)dz=1$. Example plots for different values of $r$ with $\lambda_-^{(0)}=1=2\alpha$ are shown in Fig. \ref{fig5}(b). Note that $q^*(z)$ is singular when $r<2\alpha$ but still has unit normalisation.

The embedded nature of the moment equations emerges when $\ell >0$. Setting $W_{\ell}(t)= M_1^{(\ell)}(t)+M_{-1}^{(\ell)}(t)$, $Z_{\ell}(t)= M_1^{(\ell)}(t)-M_{-1}^{(\ell)}(t)$ and $\lambda^{(\ell)}_{\pm }=\lambda^{(\ell)}_1\pm \lambda^{(\ell)}_{-1}$, equations (\ref{hier}a) become
\begin{subequations}
\label{hierg}
\begin{align}
\frac{dW_{\ell}(t)}{dt}&= \ell vZ_{\ell-1}(t)+\ell(\ell-1)DW_{\ell-2}(t) +\overline{h}(t)[\lambda^{(\ell)}_+-W_{\ell}(t^-)],\\
\frac{dZ_{\ell}(t)}{dt}&=-2\alpha Z_{\ell}(t)+ \ell vW_{\ell-1}(t)+\ell(\ell-1)DZ_{\ell-2}(t) +\overline{h}(t)[\lambda^{(\ell)}_--Z_{\ell}(t^-)],\quad \ell \geq 1.
\end{align}
\end{subequations}
Consider, for example, the case $\ell=1$ where
\begin{subequations}
\label{hier1}
\begin{align}
\frac{dW_{1}(t)}{dt}&=  vZ_{0}(t) +\overline{h}(t)[\lambda^{(1)}_+-W_1(t^-)],\\
\frac{dZ_{1}(t)}{dt}&=v-2\alpha Z_{1}(t)+\overline{h}(t)[\lambda^{(1)}_--Z_{1}(t^-)].
\end{align}
\end{subequations}
The equation for $Z_1(t)$ is closed and can be analysed along similar lines to $Z_0(t)$. On the other hand, $W_1(t)$ couples to $Z_0(t)$, which resets to $\lambda^{(0)}_-$ whenever $W_1(t)$ resets to $\lambda_+^{(1)}$. Thus the lower order resetting dynamics for $Z_0(t)$ is embedded in the resetting dynamics for $W_1(t)$. Hence, in order to determine the NESS for the stochastic variable $W_1(t)$ we would need to consider the joint probability density for the pair $Z_0(t),W_1(t)$. Similarly, for $\ell=2$,
\begin{subequations}
\label{hier2}
\begin{align}
\frac{dW_{2}(t)}{dt}&=  2vZ_{1}(t) +2D+\overline{h}(t)[\lambda^{(2)}_+-W_2(t^-)],\\
\frac{dZ_{2}(t)}{dt}&=2vW_1(t)-2\alpha Z_{2}(t)+2DZ_0(t)+\overline{h}(t)[\lambda^{(2)}_--Z_{2}(t^-)].
\end{align}
\end{subequations}
Now $Z_1(t)$ is embedded in the dynamics of $W_2(t)$ whereas both $Z_0(t)$ and $W_1(t)$ are embedded in the dynamics of $Z_2(t)$. In conclusion, as we proceed up the hierarchy, it is necessary to deal with embeddings of increasing dimensionality.

\section{Discussion}

In this paper we used the stochastic calculus of diffusion and Poisson processes to develop a general probabilistic framework for studying run-and-tumble motion in 1D at both the single-particle and population levels. We applied the theory to an RTP with position and velocity resetting and to an RTP in the presence of a partially absorbing sticky boundary. In the latter case, the probability of absorption was taken to depend on the amount of time the RTP is stuck at the boundary -- a so-called encounter-based model of absorption. We then calculated stochastic entropy production along individual trajectories of an RTP, and showed how previous expressions for the Gibbs-Shannon entropy are recovered after averaging over the ensemble of sample paths. Finally, we derived a stochastic CK equation for the population density of non-interacting RTPs with global resetting and highlighted two major consequences of the latter. First, global resetting induces statistical correlations between the non-interacting particles. Second, the moment equations of the corresponding population density form a hierarchy of ODEs with stochastic resetting in which lower-order moments are embedded into the equations at higher orders. We hope to develop a more detailed theory of this novel class of stochastic process elsewhere. Another natural generalisation of our work would be to consider higher-dimensional run-and-tumble motion \cite{Mori20,Santra20,Santra20a,Doussal22,Pal23}. Of particular relevance to the current paper would be the case where there is a discrete set of possible directions for each run, as considered in Refs. \cite{Santra20,Pal23}.  

\setcounter{equation}{0}
\renewcommand{\theequation}{A.\arabic{equation}}
\section*{Appendix A: Calculation of the NESS}

In Ref. \cite{Evans18} the NESS of the total probability $p^*(x)=p_1^*(x)+p_{-1}^*(x)$ was calculated using a renewal equation under the resetting condition $\sigma(t)\rightarrow \pm 1$ with equal probability. Since we are interested in the NESS of the individual components and do not assume a symmetric resetting rule for the velocity state, we proceed by directly solving the Laplace transform of the CK equation (\ref{CK0r}) with resetting:
\begin{align}
  vk\frac{\partial \p_k(x,s)}{\partial x}+(\alpha +s+r)\p_{k}(x,s)-\alpha \p_{-k}(x,t) =\left (1+\frac{r}{s}\right )\delta(x-x_0)\delta_{k,\sigma_0}.\label{CKLT}
\end{align}

\subsection*{\bf Case $\sigma_0=1$}  
First suppose that $\sigma_0=1$. We then have the pair of equations
\begin{align}
\label{CKLTa}
  v\frac{\partial \p_1(x,s)}{\partial x}+(\alpha +s+r)\p_{1}(x,s)-\alpha \p_{-1}(x,s) &=\left (1+\frac{r}{s}\right )\delta(x-x_0),\\
 -v\frac{\partial \p_{-1}(x,s)}{\partial x}+(\alpha +s+r)\p_{-1}(x,s)-\alpha \p_{1}(x,s) &=0.\label{CKLTb}
\end{align}
Introduce the Green's function $G_1(x,x_0)$ with
\begin{align}
 \frac{\partial G_1(x,x_0;s)}{\partial x}+\lambda(s) G_1(x,x_0;s)=\delta(x-x_0)
 \end{align}
 and
 \begin{equation}
\lambda(s)=\frac{\alpha +s+r}{v}.
 \end{equation}
Solving for $G_1$ gives
\begin{equation}
G_1(x,x_0;s)=\e^{-\lambda(s)(x-x_0)}\Theta(x-x_0),
\end{equation}
with $\Theta(x)=1$ for $x>0$ and  zero otherwise. Substituting
\begin{equation}
\label{p1LT}
\p_1(x,s)=\frac{1}{v}\left (1+\frac{r}{s}\right )G_1(x,x_0;s)+\q_1(x,s)
\end{equation}
 into equation (\ref{CKLTa}) implies that
 \begin{equation}
 \frac{\partial \q_1(x,s)}{\partial x}+\lambda(s) \q_1(x,s)=\frac{\alpha}{v}\p_{-1}(x,s).
 \label{qLT}
 \end{equation}

The next step is to substitute the solution (\ref{p1LT}) into (\ref{CKLTb}):
\begin{equation}
  -\frac{\partial \p_{-1}(x,s)}{\partial x}+\lambda(s) \p_{-1}(x,s)=\frac{\alpha}{v}\bigg \{\frac{1}{v}\left (1+\frac{r}{s}\right )G_1(x,x_0;s)+\q_1(x,s)\bigg \}
 \end{equation}
 Applying the operator $\lambda(s)+\partial/\partial x$ to both sides and rearranging, we have
 \begin{align}
 &\frac{\partial^2 \p_{-1}(x,s)}{\partial x^2}-\Lambda(s)^2\p_{-1}(x,s)
   =-\frac{\alpha}{v^2}\left (1+\frac{r}{s}\right )\delta(x-x_0),
\end{align}
where
\begin{equation}
\Lambda(s)=\frac{1}{v}\sqrt{(s+r)(2\alpha+s+r)}.
\end{equation}
Noting that the Green's function 
\begin{equation}
 G(x,x_0;s)=\frac{1}{2\Lambda(s)}\e^{-\Lambda(s) |x-x_0|}.
 \end{equation}
 is the solution (in the distribution sense) to the equation
\begin{equation}
\frac{\partial^2 G}{\partial x^2}-\Lambda^2(s)G=-\delta(x-x_0),
\end{equation}
it follows that
\begin{equation}
\p_{-1}(x,s)=\frac{\alpha}{v^2}\left (1+\frac{r}{s}\right )G(x,x_0;s).
\end{equation}

The final step is to substitute for $\p_{-1}$ in equation (\ref{qLT}) and solve for $\q_1$:
\begin{align}
\widetilde{q}_1(x,s)&=\frac{1}{2\Lambda(s)}\frac{\alpha^2}{v^3}\left (1+\frac{r}{s}\right )\int_{-\infty}^x\e^{-\lambda(s)(x-y)-\Lambda(s)|y-x_0|}dy.
\end{align}
Denote the integral on the right-hand side by ${\mathcal I}_1$. If $x>x_0$ then
\begin{align}
{\mathcal I}_1&=\e^{-\lambda x}\e^{-\Lambda x_0}\int_{-\infty}^{x_0} \e^{(\lambda+\Lambda)y}dy+\e^{-\lambda x}\e^{\Lambda x_0}\int_{x_0}^{x} \e^{(\lambda-\Lambda)y}dy\nonumber\\
&=\frac{\e^{-\lambda (x-x_0)}}{\lambda+\Lambda}+\frac{\e^{-\lambda x}\e^{\Lambda x_0}}{\lambda-\Lambda}\left [ \e^{(\lambda-\Lambda)x}- \e^{(\lambda-\Lambda)x_0}\right ]\nonumber \\
&=\frac{\e^{-\lambda (x-x_0)}}{\lambda+\Lambda}+\frac{\e^{-\Lambda (x-x_0)}}{\lambda-\Lambda}-\frac{\e^{-\lambda (x-x_0)}}{\lambda-\Lambda}=\frac{\e^{-\Lambda (x-x_0)}}{\lambda-\Lambda}- \frac{2\Lambda\e^{-\lambda (x-x_0)}}{\lambda^2-\Lambda^2}.
\end{align}
On the other hand, if $x <x_0$ then
\begin{align}
{\mathcal I}_1&=\e^{-\lambda x}\e^{-\Lambda x_0}\int_{-\infty}^{x} \e^{(\lambda+\Lambda)y}dy=\frac{\e^{-\Lambda |x-x_0|}}{\lambda+\Lambda}.
\end{align}
Combining all of our results leads to the following solution in Laplace space:
\begin{align}
 \p_1(x,s)&=\frac{1}{v}\left (1+\frac{r}{s}\right )\e^{-\lambda(s)(x-x_0)}\Theta(x-x_0)\nonumber \\
 &\quad +\frac{1}{2\Lambda(s)}\frac{\alpha^2}{v^3}\left (1+\frac{r}{s}\right )\left [\frac{\e^{-\Lambda(s) (x-x_0)}}{\lambda-\Lambda}- \frac{2\Lambda(s)\e^{-\lambda(s) (x-x_0)}}{\lambda(s)^2-\Lambda(s)^2}\right ]\Theta(x-x_0)\nonumber \\
 &\quad +\frac{1}{2\Lambda(s)}\frac{\alpha^2}{v^3}\left (1+\frac{r}{s}\right )\frac{\e^{-\Lambda(s) |x-x_0|}}{\lambda(s)+\Lambda(s)}\Theta(x_0-x),
\label{1pLT1}
\end{align}
and
\begin{equation}
\label{1pLTb}
\p_{-1}(x,s)=\frac{\alpha}{v^2}\frac{1}{2\Lambda(s)}\left (1+\frac{r}{s}\right )\e^{-\Lambda(s)|x-x_0|}.
\end{equation}

Given the solution in Laplace space, the NESS can now be written down using the identity
\begin{equation}
p_k^*(x)\equiv \lim_{t\rightarrow \infty} p_k(x,t)=\lim_{s\rightarrow 0}s\p_k(x,s).
\end{equation}
Hence, multiplying both sides of equations (\ref{1pLT1}) and (\ref{1pLTb}) by $s$ and taking $s\rightarrow 0$ gives
 \begin{align}
\left .\p_{1}^*(x)\right |_{\sigma_0=1}&=\frac{r}{v}\e^{-\lambda_r(x-x_0)}\Theta(x-x_0) +\frac{
r}{2\Lambda_r}\frac{\alpha^2}{v^3}\left [\frac{\e^{-\Lambda_r (x-x_0)}}{\lambda_r-\Lambda_r}- \frac{2\Lambda_r\e^{-\lambda_r (x-x_0)}}{\lambda_r^2-\Lambda_r^2}\right ]\Theta(x-x_0)\nonumber \\
&\quad +\frac{r}{2\Lambda_r}\frac{\alpha^2}{v^3}\frac{\e^{-\Lambda_r |x-x_0|}}{\lambda_r+\Lambda_r}\Theta(x_0-x).
\label{A1pssa}
\end{align}
and
\begin{equation}
\label{A1pssb}
\left .\p_{-1}^*(x)\right |_{\sigma_0=1}=\frac{\alpha}{v^2}\frac{r}{2\Lambda_r}\e^{-\Lambda_r|x-x_0|}.
\end{equation}
We have also set
\begin{equation}
\Lambda_r=\Lambda(0)=\frac{\sqrt{r(r+2\alpha)}}{v},\quad \lambda_r=\lambda(0)=\frac{\alpha+r}{v}.
\end{equation}
Note that $\lambda_r^2-\Lambda_r^2=\alpha^2/v^2$.

\subsection*{Case $\sigma_0=-1$}

We now follow similar steps for the CK equation with $\sigma_0=-1$: 
\begin{align}
\label{CKLTam}
  v\frac{\partial \p_1(x,s)}{\partial x}+(\alpha +s+r)\p_{1}(x,s)-\alpha \p_{-1}(x,s) &=0,\\
 -v\frac{\partial \p_{-1}(x,s)}{\partial x}+(\alpha +s+r)\p_{-1}(x,s)-\alpha \p_{1}(x,s) &=\left (1+\frac{r}{s}\right )\delta(x-x_0).\label{CKLTbm}
\end{align}
Introduce the Green's function $G_{-1}(x,x_0)$ with
\begin{align}
\frac{\partial G_{-1}(x,x_0;s)}{\partial x}-\lambda(s) G_{-1}(x,x_0;s)=-\delta(x-x_0).
 \end{align}
This has the solution
\begin{equation}
G_{-1}(x,x_0;s)=\e^{-\lambda(s)|x-x_0|}\Theta(x_0-x).
\end{equation}
Substituting
\begin{equation}
\label{p1LTm}
\p_{-1}(x,s)=\frac{1}{v}\left (1+\frac{r}{s}\right )G_{-1}(x,x_0;s)+\q_{-1}(x,s)
\end{equation}
 into equation (\ref{CKLTbm}) implies that
 \begin{equation}
 \frac{\partial \q_{-1}(x,s)}{\partial x}-\lambda(s) \q_1(x,s)=-\frac{\alpha}{v}\p_{1}(x,s).
 \label{qLTm}
 \end{equation}
Substitute the solution (\ref{p1LTm}) into (\ref{CKLTam}):
\begin{equation}
  \frac{\partial \p_{1}(x,s)}{\partial x}+\lambda(s) \p_{1}(x,s)=\frac{\alpha}{v}\bigg \{\frac{1}{v}\left (1+\frac{r}{s}\right )G_{-1}(x,x_0;s)+\q_{-1}(x,s)\bigg \}
 \end{equation}
 Applying the operator $\partial/\partial x-\lambda(s)$ to both sides and rearranging, we have
 \begin{align}
 &\frac{\partial^2 \p_{1}(x,s)}{\partial x^2}-\Lambda(s)^2\p_{1}(x,s)
   =-\frac{\alpha}{v^2}\left (1+\frac{r}{s}\right )\delta(x-x_0),
\end{align}
and thus
\begin{equation}
\p_{1}(x,s)=\frac{\alpha}{v^2}\left (1+\frac{r}{s}\right )G(x,x_0;s).
\end{equation}
Next, we substitute for $\p_{1}$ in equation (\ref{qLTm}) and solve for $\q_{-1}$:
\begin{align}
\widetilde{q}_{-1}(x,s)&=\frac{1}{2\Lambda(s)}\frac{\alpha^2}{v^3}\left (1+\frac{r}{s}\right )\int_x^{\infty}\e^{\lambda(s)(x-y)-\Lambda(s)|y-x_0|}dy.
\end{align}
Denote the integral on the right-hand side by ${\mathcal I}_{-1}$. If $x>x_0$ then
\begin{align}
{\mathcal I}_{-1}&=\e^{\lambda x}\e^{\Lambda x_0}\int_x^{\infty} \e^{-(\lambda+\Lambda)y}dy=\frac{\e^{-\Lambda (x-x_0)}}{\lambda+\Lambda}.
\end{align}
On the other hand, if $x <x_0$ then
\begin{align}
{\mathcal I}_{-1}&=\e^{\lambda x}\e^{-\Lambda x_0}\int_{x}^{x_0} \e^{-(\lambda-\Lambda)y}dy+\e^{\lambda x}\e^{\Lambda x_0}\int_{x_0}^{\infty} \e^{-(\lambda+\Lambda)y}dy\nonumber\\
&=\frac{\e^{\lambda x}\e^{-\Lambda x_0}}{\lambda-\Lambda}\left [ \e^{-(\lambda-\Lambda)x}- \e^{-(\lambda-\Lambda)x_0}\right ]+\frac{\e^{-\lambda |x-x_0|}}{\lambda+\Lambda}\nonumber \\
&=\frac{\e^{-\Lambda |x-x_0|}}{\lambda-\Lambda}-\frac{\e^{-\lambda |x-x_0|}}{\lambda-\Lambda}+\frac{\e^{-\lambda |x-x_0|}}{\lambda+\Lambda}=\frac{\e^{-\Lambda |x-x_0|}}{\lambda-\Lambda}- \frac{2\Lambda\e^{-\lambda |x-x_0|}}{\lambda^2-\Lambda^2}.
\end{align}

Combining all of our results leads to the following solution in Laplace space:
\begin{equation}
\label{-1pLTa}
\p_{1}(x,s)=\frac{\alpha}{v^2}\frac{1}{2\Lambda(s)}\left (1+\frac{r}{s}\right )\e^{-\Lambda(s)|x-x_0|} .
\end{equation}
and
\begin{align}
 \p_{-1}(x,s)&=\frac{1}{v}\left (1+\frac{r}{s}\right )\e^{-\lambda(s)|x-x_0|}\Theta(x_0-x)\nonumber \\
 &\quad +\frac{1}{2\Lambda(s)}\frac{\alpha^2}{v^3}\left (1+\frac{r}{s}\right )\left [\frac{\e^{-\Lambda(s) |x-x_0|}}{\lambda(s)-\Lambda(s)}-\frac{2\Lambda(s)\e^{-\lambda(s) |x-x_0|}}{\lambda(s)^2-\Lambda(s)^2}\right ]\Theta(x_0-x)\nonumber \\
 &\quad +\frac{1}{2\Lambda(s)}\frac{\alpha^2}{v^3}\left (1+\frac{r}{s}\right )\frac{\e^{-\Lambda(s) |x-x_0|}}{\lambda(s)+\Lambda(s)}\Theta(x-x_0),
\label{-1pLTb}
\end{align}
Finally, we obtain the NESS for $\sigma_0=-1$:
\begin{equation}
\label{A-1pssa}
\left .\p_{1}^*(x)\right |_{
\sigma_0=-1}=\frac{\alpha}{v^2}\frac{r}{2\Lambda_r}\e^{-\Lambda_r|x-x_0|}.
\end{equation}
and
 \begin{align}
\left .\p_{-1}^*(x)\right |_{\sigma_0=-1}&=\frac{r}{v}\e^{-\lambda_r|x-x_0|}\Theta(x_0-x )
 +\frac{r}{2\Lambda_r}\frac{\alpha^2}{v^3}\left [\frac{\e^{-\Lambda_r |x-x_0|}}{\lambda_r-\Lambda_r}-\frac{2\Lambda_r\e^{-\lambda_r |x-x_0|}}{\lambda_r^2-\Lambda_r^2}\right ]\Theta(x_0-x)\nonumber \\
&\quad +\frac{r}{2\Lambda_r}\frac{\alpha^2}{v^3}\frac{\e^{-\Lambda_r |x-x_0|}}{\lambda_r+\Lambda_r}\Theta(x-x_0).
\label{A-1pssb}
\end{align}

\subsection*{Symmetric case} A useful check of the above calculations is to compare them with the expression  for the total probability density $p^*(x)=p_1^*(x)+p_{-1}^*(x)$ obtained in Ref. \cite{Evans18} for the symmetric resetting condition $\sigma_0=\pm 1$ with equal probability. In terms of the above results, we have
\begin{align}
 p^*(x)&=\frac{1}{2}\left [\left .\p_{1}^*(x)\right |_{
\sigma_0=1}+\left .\p_{1}^*(x)\right |_{
\sigma_0=-1}+\left .\p_{-1}^*(x)\right |_{
\sigma_0=1}+\left .\p_{-1}^*(x)\right |_{
\sigma_0=-1}\right ]\nonumber \\
 &=\frac{\alpha}{v^2}\frac{r}{2\Lambda_r}\e^{-\Lambda_r|x-x_0|}+\frac{r}{2v}\e^{-\lambda_r|x-x_0|} \nonumber  \\
 &\quad +\frac{r}{4\Lambda_r}\frac{\alpha^2}{v^3}\left [\frac{\e^{-\Lambda_r (x-x_0)}}{\lambda_r-\Lambda_r}- \frac{2\Lambda_r\e^{-\lambda_r (x-x_0)}}{\lambda_r^2-\Lambda_r^2}\right ] +\frac{r}{4\Lambda_r}\frac{\alpha^2}{v^3}\frac{\e^{-\Lambda_r |x-x_0|}}{\lambda_r+\Lambda_r} \nonumber \\
 & =\frac{r}{2v\Lambda_r}\left [\frac{\alpha}{v}+\lambda\right ]\e^{-\Lambda_r|x-x_0|} 
=\frac{\Lambda_r}{2}\e^{-\Lambda_r|x-x_0|}.
\end{align}
This recovers of the NESS of Ref. \cite{Evans18}.


\begin{thebibliography}{9}

\bibitem{Ramaswamy10} Ramaswamy S. 2010 The mechanics and statistics
of active matter. {\em Annu. Rev. Condens. Matter Phys.} {\bf 1}, 323-345 

\bibitem{Palacci10} Palacci J, Cottin-Bizonne C, Ybert C, Bocquet L. 2010 Sedimentation and effective temperature of active colloidal suspensions. {\em Phys. Rev. Lett.} {\bf 105}, 088304



\bibitem{Vicsek12} Vicsek T, Zafeiris A. 2012 Collective motion. {\em Phys. Rep.}
{\bf 517}, 71-140 (2012).


\bibitem{Roman12} Romanczuk R, Bar M, Ebeling W, Lindner B,
Schimansky-Geier L. 2012 Active Brownian particles:
From individual to collective stochastic dynamics. {\em Eur. Phys. J. Special Topics }{\bf 202}, 1-162  

\bibitem{Bricard13} Bricard A, Caussin JB, Desreumaux N, Dauchot O, Bartolo D. 2013 Emergence of macroscopic directed motion in populations of motile colloids. {\em Nature} {\bf 503}, 95


\bibitem{Solon15}  Solon AP, Cates ME, Tailleur J. 2015
Active brownian particles and run-and-tumble particles: A comparative study.
{\em Eur. Phys. J. Special Topics} {\bf 224}, 1231-1262 
 
 
\bibitem{Attanasi14} Attanasi A et al. 2014 Information transfer and behavioural inertia in starling flocks. {\em Nat. Phys.} {\bf 10}, 691

\bibitem{Cates15} Cates ME, Tailleur J. 2015 Motility-induced phase
separation. {\em Annu. Rev. Condens. Matter Phys.} {\bf 6}, 219-244 

\bibitem{Bechinger16} Bechinger C, Di Leonardo R, Lowen H, Reichhardt C, Volpe G, Volpe G. 2016 Active particles in complex and crowded environments. {\em Rev. Mod. Phys.} {\bf 88}, 045006

  \bibitem{Berg04} Berg HC. 2004 {\em E. Coli in Motion}, New York, Springer

\bibitem{Angelani15} Angelani L. 2015 Run-and-tumble particles, telegrapher's equation
and absorption problems with partially reflecting boundaries
{\em J. Phys. A: Math. Theor.} {\bf 48}, 495003 

\bibitem{Angelani17} Angelani L. 2017 Confined run-and-tumble swimmers in one dimension. {\em J. Phys. A} {\bf 50}, 325601  
 
 \bibitem{Malakar18} Malakar K, Jemseena V, Kundu A, Vijay Kumar, 
Sabhapandit S, Majumdar SN, Redner S and Dhar A. 2018 Steady state, relaxation and first-passage properties of a run-and-tumble particle in one-dimension, {\em J. Stat.
Mech.} 043215 

\bibitem{Bressloff22d} Bressloff PC. 2022 Encounter-based model of a run-and-tumble particle. {\em J. Stat. Mech.} 113206 (2022)



\bibitem{Bressloff23} Bressloff PC. 2023 Encounter-based model of a run-and-tumble particle II: absorption at sticky boundaries. {\em J. Stat. Mech.} {\bf 043208}


 \bibitem{Angelani23} Angelani L. 2023  One-dimensional run-and-tumble motions
with generic boundary conditions. {\em J. Phys. A} {\bf 56} ,455003.


  \bibitem{Evans18} Evans MR, Majumdar SN. 2018 Run and tumble particle under resetting: a renewal approach. {\em J. Phys. A: Math. Theor.} {\bf 51} 475003, (2018).


\bibitem{Bressloff20} Bressloff PC. 2020 Occupation time of a run-and-tumble particle with resetting. {\em Phys. Rev. E} {\bf 102} 042135 

\bibitem{Santra20a} Santra I, Basu U, Sabhapandit S> 2020 Run-and-tumble particles in two dimensions under stochastic resetting conditions {\em J. Stat. Mech.} 113206

\bibitem{Grebenkov20} Grebenkov DS. 2020  {Paradigm shift in diffusion-mediated surface phenomena.} {\em Phys. Rev. Lett.} {\bf 125} 078102  


\bibitem{Grebenkov22} Grebenkov DS. 2022  {An encounter-based approach for restricted diffusion with a gradient drift.}  {\em J. Phys. A.} {\bf 55} 045203 



\bibitem{Bressloff22} Bressloff PC. 2022  Diffusion-mediated absorption by partially reactive targets: Brownian functionals and generalised propagators. {\em J. Phys. A.} {\bf 55} 205001

\bibitem{Bressloff22a} Bressloff PC 2022 Spectral theory of diffusion in partially absorbing media. {\em Proc. R. Soc. A} {\bf 478}, 20220319

\bibitem{Bressloff22b} Bressloff PC. 2022  Diffusion-mediated surface reactions and stochastic resetting. {\em J. Phys. A} {\bf 55}, 275002 

\bibitem{Bressloff22c} Bressloff PC. 2022  Diffusion in a partially absorbing medium with position and occupation time resetting. {\em J. Stat. Mech.} {\bf 063207}.

\bibitem{Seifert05} Seifert U. 2005 Entropy production along a stochastic
trajectory and an integral fluctuation theorem. {\em Phys. Rev. Lett.} {\bf 95}, 040602 


\bibitem{Sekimoto10} Sekimoto K. 2010 {\em Stochastic Energetics.} Lecture Notes in Physics, Springer

\bibitem{Seifert12} Seifert U. 2012 Stochastic thermodynamics, fluctuation theorems
and molecular machines. {Rep. Prog. Phys.} {\bf 75} ,126001 

\bibitem{Cocconi20} Cocconi, L Garcia-Millan R, Zhen Z, Buturca B, Pruessner G. 2020  Entropy production in exactly solvable systems. {\em Entropy} {\bf 22}, 1252  

 \bibitem{Peliti21} Peliti L, Pigolotti S. 2021 {\em Stochastic Thermodynamics}. Princeton University Press, Princeton.
 

\bibitem{Roldan23} R\'oldan E, Neri I, Chetrite R, Pigolotti S, J\"ulicher F,  
Sekimoto K. 2023 Martingales for physicists:
a treatise on stochastic thermodynamics and beyond {\em Advances in Physics} {\bf 72}, 2317494

\bibitem{Frydel22} Frydel D. 2022 Intuitive view of entropy production of ideal run-and-tumble particles. {\em Phys. Rev. E} {\bf 105}, 034113


\bibitem{Angelani24} Paoluzzi M, Puglisi A, Angelani L. 2024 Entropy production of
run-and-tumble particles. {\em Entropy}
{\bf 26}, 443

\bibitem{Fuchs16} Fuchs J, Goldt S, Seifert U. 2016. Stochastic thermodynamics
of resetting {\em Europhys. Lett.} {\bf 113}, 60009  

\bibitem{Mori23} Mori F, Olsen KS, Krishnamurthy S. 2023 Entropy production of resetting processes, {\em Phys. Rev. Res.}, {\bf 5}, 123103  




\bibitem{Bressloff24a} Bressloff PC. 2024 Global density equations for interacting particle systems with stochastic resetting: from overdamped Brownian motion to phase synchronization {\em Chaos} {\bf 34}, 043101.  


\bibitem{Bressloff24b} Bressloff PC. 2024 A generalised Dean-Kawasaki equation for an interacting Brownian gas in a partially absorbing medium. {\em Proc. Roy. Soc.} {\bf 480},
20230915.

 \bibitem{Nagar23} Nagar A, Gupta S. 2023 Stochastic resetting in interacting particle
systems: a review' {\em J. Phys. A: Math. Theor.} {\bf 56}, 283001.




 \bibitem{Mori20} Mori F, Le Doussal P, Majumdar SN, Schehr G. 2020 Universal Survival Probability for a d-Dimensional Run-and-Tumble Particle. {\em Phys. Rev. Lett.} {\bf 124}, 090603
  
\bibitem{Santra20} Santra I, Basu U, Sabhapandit S. 2020 Run-and-tumble particles in two-dimensions: Marginal position distributions {\em Phys. Rev. E} {\bf 101} 062120. 
 


\bibitem{Doussal22} Smith NR, Le Doussal P, Majumdar, SN, Schehr G. 2022 Exact position distribution of a harmonically confined run-and-tumble particle in two dimensions {\em Phys. Rev. E} {\bf 106}, 054133

\bibitem{Pal23} Mallikarjun R, Pal A. 2023 Chiral run-and-tumble walker: Transport and optimizing
search {\em Physica A} {\bf 622}, 128821

\end{thebibliography}
\end{document}